\documentclass[aoas,MSNbibl,nameyear,dvips]{arximspdf}
\usepackage{multirow,dcolumn}
\usepackage{graphicx}
\usepackage{url,breakurl}

\doi{10.1214/14-AOAS775} 
\volume{8}
\issue{4}
\pubyear{2014}
\firstpage{2378}
\lastpage{2403}
\docsubty{FLA}

\makeatletter
\newcolumntype{d}[1]{D{.}{.}{#1}}
\makeatother

\begin{document}
\begin{frontmatter}

\title{Synthesising evidence to estimate pandemic (2009) A/H1N1
influenza severity in 2009--2011\thanksref{T1}}
\runtitle{Synthesising evidence to estimate influenza severity}
\thankstext{T1}{Supported in part by the Medical Research
Council [Unit Programme Numbers U105260566 and U105261167], NIHR
HTA Project 11/46/03, Public Health England and the Royal
College of General Practitioners.}
\begin{aug}
\author[A]{\fnms{Anne~M.} \snm{Presanis}\corref{}\ead[label=e1]{anne.presanis@mrc-bsu.cam.ac.uk}\thanksref{m1}},
\author[B]{\fnms{Richard~G.} \snm{Pebody}\thanksref{m2}},
\author[A]{\fnms{Paul~J.} \snm{Birrell}\thanksref{m1}},
\author[A]{\fnms{Brian~D.~M.} \snm{Tom}\thanksref{m1}},
\author[B]{\fnms{Helen~K.} \snm{Green}\thanksref{m2}},
\author[C]{\fnms{Hayley} \snm{Durnall}\thanksref{m3}},
\author[C]{\fnms{Douglas} \snm{Fleming}\thanksref{m3}}
\and
\author[A]{\fnms{Daniela} \snm{De Angelis}\thanksref{m1}}
\runauthor{Presanis et al.}
\affiliation{Medical Research Council Biostatistics Unit\thanksmark
{m1},
Public Health England\thanksmark{m2}
and Royal College of General Practitioners\thanksmark{m3}}

\address[A]{A. M. Presanis\\
P. J. Birrell\\
B. D. M. Tom\\
D. De Angelis\\
Medical Research Council\\
\quad Biostatistics Unit\\
Cambridge Institute of Public Health\\
Forvie Site Robinson Way\\
Cambridge Biomedical Campus\\
Cambridge CB2 0SR\\
United Kingdom\\
\printead{e1}}
\address[B]{R. G. Pebody\\
H. K. Green\\
Public Health England\\
61 Colindale Avenue\\
London NW9 5EQ\\
United Kingdom}

\address[C]{H. Durnall\\
D. Fleming\\
RCGP Research \& Surveillance Centre\\
Lordswood Medical Group\\
54 Lordswood Road Harborne\\
Birmingham B17 9DB\\
United Kingdom}

\end{aug}
%

\received{\smonth{7} \syear{2013}}
\revised{\smonth{2} \syear{2014}}

%
\begin{abstract}
Knowledge of the severity of an influenza outbreak is crucial for
informing and monitoring appropriate public health responses, both
during and after an epidemic. However, case-fatality, case-intensive
care admission and case-hospitalisation risks are difficult to measure
directly. Bayesian evidence synthesis methods have previously been
employed to combine fragmented, under-ascertained and biased
surveillance data coherently and consistently, to estimate
case-severity risks in the first two waves of the 2009 A/H1N1 influenza
pandemic experienced in England. We present in detail the complex
probabilistic model underlying this evidence synthesis, and extend the
analysis to also estimate severity in the third wave of the pandemic
strain during the 2010/2011 influenza season. We adapt the model to
account for changes in the surveillance data available over the three
waves. We consider two approaches: (a) a two-stage approach using
posterior distributions from the model for the first two waves to
inform priors for the third wave model; and (b) a one-stage approach
modelling all three waves simultaneously. Both approaches result in the
same key conclusions: (1) that the age-distribution of the
case-severity risks is ``u''-shaped, with children and older adults
having the highest severity; (2) that the age-distribution of the
infection attack rate changes over waves, school-age children being
most affected in the first two waves and the attack rate in adults over
25 increasing from the second to third waves; and (3) that when
averaged over all age groups, case-severity appears to increase over
the three waves. The extent to which the final conclusion is driven by
the change in age-distribution of those infected over time is subject
to discussion.
\end{abstract}

%
\end{frontmatter}

\section{Introduction}
Evidence synthesis [e.g., Spiegelhalter, Abrams and\break Myles
(\citeyear{SpiegelhalterEtAl2004}), \citet{AdesSutton2006}]
has become an important method in epidemiology, where multiple,
disparate, incomplete and often biased sources of observational (e.g.,
surveillance or survey) data are available to inform estimation of
relevant quantities, such as prevalence and incidence of infectious
disease [\citet{WeltonAdes2005,GoubarEtAl2008,SweetingEtAl2008,AlbertEtAl2011,PresanisEtAl2011,BirrellEtAl2011}].
Data may directly inform a quantity of interest, $\theta$, or, more
usually, may indirectly inform multiple parameters $\bolds{\Theta
}$ by directly informing some function of $\bolds{\Theta}$, $\psi
= \psi(\bolds{\Theta})$. Such a function may represent, for
example, the relationship between a biased source of data and the
parameter the data should theoretically measure, so that the bias is
explicitly modelled. Evidence synthesis methods combine these
heterogeneous types of challenging data in a coherent manner, to
estimate the ``basic'' parameters $\bolds{\Theta}$ and from these
obtain simultaneously the ``functional'' parameters $\bolds{\Psi}
= \{\psi_1(\bolds{\Theta}), \ldots, \psi_m(\bolds{\Theta})\}
$. These functional parameters include both those directly observed and
others that may not be observed but are of interest to estimate. This
type of estimation typically necessitates the formulation of complex
probabilistic models, often in a Bayesian framework.

Knowledge of the severity of an influenza outbreak is crucial for
informing and monitoring appropriate public health responses. Severity
estimates are necessary not only during a pandemic to inform immediate
public health responses, but also afterwards, when a robust
reconstruction of what happened during the pandemic is required to
evaluate the responses. Moreover, as has happened in past influenza
pandemics [\citet{MillerEtAl2009}], if a pandemic strain continues to
circulate for some years, with unusual patterns of age-specific
mortality, then severity estimates over time, both in terms of attack
rates (the proportion of the population infected) and case-severity
risks (the probability an infection leads to a severe event), are
required to understand if the strain is likely to continue circulating
and if severity is changing over time.

However, severity is an example epidemic characteristic that is
difficult to measure directly. Typically, severity is expressed as the
probability that an infection will result in a severe event, for
example, death. We refer to this probability as the ``case-fatality
risk'' ($\mathit{CFR}$). Severity may also be quantified by
``case-hospitalisation'' ($\mathit{CHR}$) and ``case-intensive care admission''
($\mathit{CIR}$) risks, defined similarly as probabilities that an infection
results in hospitalisation or intensive care (ICU) admission. Not all
influenza infections will be symptomatic, where ``symptomatic'' may be
defined in different ways, but is here taken to denote febrile
influenza-like illness (ILI). Not all infections will therefore result
in symptoms severe enough for a patient to access health care and hence
be detectable in surveillance systems
[\citet{ReedEtAl2009,PresanisEtAl2011a,BirrellEtAl2011}]. Symptomatic
case-severity risks ($\mathit{sCHR}, \mathit{sCIR}, \mathit{sCFR}$),
the probabilities a
symptomatic infection leads to severe events, are therefore also
considered as important indicators of severity for influenza.
Estimation of these probabilities requires information on both the
cumulative incidence of (symptomatic) infection over a period of time
of interest (the denominator) and the cumulative incidence of severe
events (the numerator). However, the denominator,\vadjust{\goodbreak} whether symptomatic
or all infection, is challenging to determine, due to the unobserved
infections. Population-wide serological testing (testing for antibodies
to influenza infection in blood serum samples) to measure the
proportion of the population infected is one possibility, but is
unlikely to be feasible. This challenge is only compounded in a
pandemic situation, where resources and time are even more stretched
than usual [e.g., \citet{LipsitchEtAl2009,GarskeEtAl2009}].

The most feasible approach to the assessment of severity is therefore
via estimation, combining data from different sources and accounting
for their biases, due, for example, to under-ascertainment. The
majority of methods adopted to estimate influenza case-severity [e.g.,
\citet{ReedEtAl2009,GarskeEtAl2009,WilsonBaker2009,PebodyEtAl2010,WieldersEtAl2010,SypsaEtAl2011}]
have not systematically accounted for all biases. Crucially, they have
not made use of all available information in the estimation process,
nor have they accounted for all uncertainty inherent in the data.
Bayesian evidence synthesis provides a flexible framework in which all
available relevant data may be coherently amalgamated, together with
prior information on biases, to estimate case-severity
[\citet{LipsitchEtAl2011,McDonaldEtAl2013},
\citeauthor{PresanisEtAl2009} (\citeyear{PresanisEtAl2009,PresanisEtAl2011a}),
\citet{ShubinEtAl2013,WuEtAl2010}].

Until the 2012/2013 winter, England experienced three waves of infection
with the 2009 pandemic A/H1N1 influenza strain: in the summer of 2009,
the autumn and winter of 2009--2010, and the autumn and winter of 2010--2011.
The severity of the first two waves, as measured by case-severity
risks, was previously estimated [\citet{PresanisEtAl2011a}] by
synthesising data either from surveillance systems in place to monitor
seasonal influenza or from systems set up specifically in response to
the pandemic [\citet{HPAepiReport2010}]. In this paper, we present in
\mbox{detail} the statistical model used in \citet{PresanisEtAl2011a} and
extend the approach to estimating severity in the third wave of
infection. After the first two waves, the World Health Organization
declared a move to a post-pandemic period (\url
{http://www.who.int/mediacentre/news/statements/2010/h1n1_vpc_20100810/en/index.html}),
at which time many of the\break surveillance systems that operated during the
pandemic situation were either stopped or changed in form. We describe
how the model of \citet{PresanisEtAl2011a} is further developed to
account for these changes in the available data.

The evidence used to estimate severity in the first two waves and the
changes to the surveillance systems between waves are described in
Section~\ref{sec_data}. A Bayesian approach to evidence synthesis is
introduced in Section~\ref{sec_methods}. We then describe in Section~\ref{sec_model} a generic model for estimating severity, before showing
in Section~\ref{sec_model12} how the model was implemented in the first
two waves. We next develop the model to estimate severity in the third
wave, presenting two approaches (Sections~\ref{sec_3wsep}
and~\ref{sec_3wsim}, resp.). Results are given in Section~\ref{sec_results}
and we end with a discussion in Section~\ref{sec_discuss}.\

\section{Surveillance data} \label{sec_data}

\subsection{First \& second waves}

During the first two pandemic waves in 2009--2010, data were available
from various surveillance systems at or used by the UK's Health
Protection Agency (HPA, now Public Health England) that provided
evidence on some aspect of the pandemic, at various levels of severity.
These sources indirectly informed the case-severity risks and full
details of each are given in Section~1.1 of the supplementary material
[\citet{PresanisEtAlSupp2013}]. Briefly, they included the following:
\begin{longlist}[(iii)]
\item[(i)] data on laboratory-confirmed pandemic A/H1N1 cases [i.e.,
cases where infection with the pandemic strain was confirmed
virologically, via real-time polymerase chain reaction (RT-PCR) testing
of nasal or throat swabs] in the first few weeks of the pandemic 
[Health Protection Agency,
Health Protection Scotland,
Communicable Disease Surveillance Centre\break Northern Ireland and
National Public Health Service for Wales (\citeyear{FF100project2009}), 
\citet{HPAepiReport2010}]. The data included dates of
illness onset and information on hospital admission if it occurred,
from which age group-specific case-hospitalisation risks amongst
confirmed cases could be estimated. Note that these
confirmed-case-hospitalisation risks are likely to be higher than the
case-hospitalisation risks in all symptomatic cases, since not all
symptomatic cases will have been confirmed in the first few weeks, and
more severe cases in hospital are more likely to have been detected
than less severe cases;
\item[(ii)] estimates of the number of symptomatic cases by week, age
and region, produced by the HPA. These estimates were recognised to be
under-estimates, given the data of point {(iii)};
\item[(iii)] serial data on age group-specific proportions of
individuals with antibodies to the pandemic strain of influenza
(``sero-prevalence''), from repeated cross-sectional surveys of
residual sera from other (unrelated) diagnostic testing [\citet{MillerEtAl2010,HardelidEtAl2011}]. These data indirectly inform the
cumulative incidence of infection, that is, the proportion of the
population infected over a period of time. Initially these data were
taken at face value, but concerns about potential sampling biases led
to extra sensitivity analyses (see Section~\ref{sec_sens12});
\item[(iv)] data on laboratory-confirmed cases in hospital [Campbell et~al.\break
(\citeyear
{CampbellEtAl2011})], including age group and dates of illness onset,
hospital admission and ICU admission; and
\item[(v)] data on the number of deaths amongst persons with confirmed
pandemic A/H1N1 influenza and/or mention of influenza on the death
certificate, reported to the HPA and/or the Chief Medical Officer [\citet{DonaldsonEtAl2009,PebodyEtAl2010}].
\end{longlist}

\subsection{Third wave}

During the third wave, data sources {(i)}, {(ii)} and
{(iv)} were no longer available in the same form. Although
results from testing of samples from before and after the third wave
from data source {(iii)} are now available [\citet{HoschlerEtAl2012}], at the time of the analyses presented here, they
were not accessible. Full details of each source below are given in
Section~1.2 of the supplementary material [\citet{PresanisEtAlSupp2013}].

\begin{longlist}[(vii)]
\item[(vi)] Between the second and third waves, the surveillance system
for hospital admissions of confirmed cases moved to being a sentinel
surveillance system, the UK Severe Influenza Surveillance Scheme
(USISS). The data from this system are available at a coarser level of
age aggregation and come from a sentinel sample of 23 acute NHS
hospital trusts in the 2010--2011 season, as opposed to the 129 trusts
participating in hospital surveillance during the first two waves.
\item[(vii)] Additional data are available on patients present in all
ICUs in England with \emph{suspected} pandemic A/H1N1 influenza, again
at a coarser age aggregation, from the Department of Health [DH; \citet{DHWW2011}].
\item[(viii)] We also have data on virological positivity (proportion
testing positive for the pandemic strain) from a sentinel system,
``Datamart,'' comprising results of RT-PCR testing from 16 HPA and NHS
laboratories in England, covering mainly patients hospitalised with
respiratory illness.
\item[(ix)] In the third wave, the HPA estimates of source
{(ii)} were not available, due to the underlying data being specified
at a different level of disaggregation. Instead, we use estimates of
the number symptomatic (details in Section~3.1 of the supplementary
material [\citet{PresanisEtAlSupp2013}]) obtained from an alternative
general practice sentinel surveillance system [\citet{Fleming1999}].
\end{longlist}

\subsection{Challenges}

Estimating case-severity by dividing the observed number of infections
at a severe level over a period of time by the observed (i.e.,
confirmed) number of infections in the same period is highly likely to
result in biased estimates. This bias is due to both
under-ascertainment of infections in surveillance systems and
differential probabilities of observation by severity of infection
[\citet{GarskeEtAl2009,PresanisEtAl2011a}]. Any estimation therefore has
to account for these probabilities of observing infections (``detection
probabilities''). Further challenges are posed by the following:
uncertainty about the representativeness of the surveillance data for
the general population (sampling biases); the different degrees of
aggregation in each data source; the fact that some of the data
sources, such as the sero-prevalence data, only inform \emph
{indirectly} the number of infections; and the changes in surveillance
systems over time. A synthesis of all the above data sources to
estimate case-severity therefore requires these challenges to be addressed.

\section{Evidence synthesis methods} \label{sec_methods}

Evidence synthesis [see, e.g., \citet{EddyEtAl1992,AdesSutton2006}]
denotes the idea of estimating a set of $k$ ``basic'' parameters
$\bolds{\Theta} = (\theta_1,\theta_2,\ldots,\theta_k)$ from a
collection of $n$ independent data sources $\mathbf{y} =
(y_1,y_2,\ldots,y_n)$, arising from multiple studies, perhaps of
differing design. Each source $y_i, i \in1,\ldots,n$ provides evidence
on a ``functional'' parameter $\psi_i = f_i(\bolds{\Theta})$. The
function $f_i$ may either be equality to a single specific element
$\theta_j$ of $\bolds{\Theta}$, so that the data \emph{directly}
informs $\theta_j$, or a function of one or more components of
$\bolds{\Theta}$, so that the data \emph{indirectly} inform
multiple basic parameters. The collection $(\psi_1,\psi_2,\ldots,\psi
_n)$ is therefore a mixture of basic and functional parameters. The aim
is to estimate the set of basic parameters $\bolds{\Theta}$, from
which the functional parameters $(\psi_1,\psi_2,\ldots,\psi_n)$, as
well as any other functions $(\psi_{n+1},\ldots,\psi_m)$ of $\bolds{\Theta}$ that are of interest, may be simultaneously derived. Denote
the total set of functions by $\bolds{\Psi}$.

Inference may be carried out either in a classical setting, maximising
the likelihood $L(\mathbf{y} | \bolds{\Theta}) = \prod_{i =
1}^n L_i(y_i | \bolds{\Theta})$, or, as in this paper, in a
Bayesian setting, assigning a prior distribution to the basic
parameters, $P(\bolds{\Theta})$, and obtaining the posterior
distribution $P(\bolds{\Theta} | \mathbf{y}) \propto
P(\bolds{\Theta}) L(\mathbf{y} | \bolds{\Theta})$
typically via a simulation-based algorithm such as Markov chain Monte
Carlo (MCMC). The posterior distribution of any of the functional
parameters may also be derived.

A Bayesian evidence synthesis meets the challenges of case-severity
estimation by allowing the relationship between data and parameters to
be accurately formulated, for example, through the use of bias
parameters such as detection probabilities; prior information on such
biases to be easily introduced; and a natural framework in which to
assess the consistency of evidence [\citet{PresanisEtAl2013}], as part
of the inference and model criticism cycle advocated by \citet{Box1980}
and \citet{OHagan2003}.

\section{A general Bayesian model for severity} \label{sec_model}

The following generic synthesis of evidence to estimate severity was
the basis of the estimation of severity of the 2009 pandemic A/H1N1
strain of influenza [\citeauthor{PresanisEtAl2009}
(\citeyear{PresanisEtAl2009,PresanisEtAl2011a})], both
in the USA and in England during the first two waves.

Assume the population of interest is divided into 7 age groups:
$<1$, 1--4, 5--14, 15--24, 25--44, 45--64, 65$+$, indexed by $a \in1,\ldots,7$.
Denote the age-specific population sizes by $N_{w,a}$, where $w$
indexes waves of infection ($w = 1, 2, 3$ in the case of England).
Consider infections at five increasing severity levels: all infections
($\mathit{Inf}$), symptomatic infections ($S$), hospitalisations ($H$), ICU
admissions ($I$) and deaths ($D$). For each wave and age-group,
consider each of these sets of infections to be subsets of the set of
infections at a less severe level, such that $D \subseteq H$ and $I
\subseteq H \subseteq S \subseteq \mathit{Inf}$. Note that we assume the set of
deaths is a subset of the set of hospitalisations, but that not all
deaths are a subset of the set of ICU admissions. The set of infections
$\mathit{Inf}$ is clearly a subset of the population.
For each age group $a$, denote the cumulative number of new infections
during wave $w$ at severity level $l$ (i.e., the size of subset $l$) by
$N_{w,a,l}$.

\subsection{Parameterisation}
Denote by $c_{w,a,l | \lambda}$ the age- and wave-specific conditional
probability that a case is at severity level $l$ given the case has
already reached a less severe level $\lambda$, that is, $l \subseteq
\lambda$. For $l = S, H, I$, let $N_{w,a,l} = c_{w,a,l | \lambda}
\times N_{w,a,\lambda}$, where $\lambda= \mathit{Inf}, S, H$, respectively. For
all infections, define $N_{w,a,\mathit{Inf}} = c_{w,a,\mathit{Inf} | \mathit{Pop}} \times
N_{w,a}$. For deaths, define $N_{w,a,D} = c_{w,a, D | H} \times
N_{w,a,H}$, that is, in terms of the conditional probability of dying
given hospitalisation. The conditional probabilities $c_{w,a, \mathit{Inf} |
\mathit{Pop}}, c_{w,a, S | \mathit{Inf}}, c_{w,a, H | S}$, $c_{w,a, I | H}$ and $c_{w,a,
D | H}$ are basic parameters to which we assign prior distributions and
the $N_{w,a,l}$ are functional parameters. Note that in the US analysis
[\citet{PresanisEtAl2009}], the $N_{w,a,l}$ were considered stochastic
nodes, realisations of a Binomial distribution with probability
parameter $c_{w,a,l | \lambda}$ and an appropriate denominator
$N_{w,a,\lambda}$. However, in the UK analysis [\citet{PresanisEtAl2011a}] and the analyses reported below, convergence of
the MCMC algorithm was only achieved when the corresponding
deterministic (mean) assumption was made for the $N_{w,a,l}$, for
reasons that are discussed further in Section~\ref{sec_discuss}.

The subsetting assumptions allow the case-hospitalisation, case-ICU
admission and case-fatality risks to be defined as functional
parameters expressed as products of component conditional probabilities:
%
\begin{eqnarray}\label{eqn_CSR}
\mathit{CHR}_{w,a} &=& c_{w,a,H | \mathit{Inf}} =  c_{w,a, H | S} \times
c_{w,a, S |
\mathit{Inf}},
\nonumber
\\
\mathit{CIR}_{w,a} &=& c_{w,a,I | \mathit{Inf}} =  c_{w,a, I | H} \times
c_{w,a, H | S} \times c_{w,a, S | \mathit{Inf}},
\\
\mathit{CFR}_{w,a} &=& c_{w,a,D | \mathit{Inf}} =  c_{w,a, D | H} \times
c_{w,a, H | S} \times c_{w,a, S | \mathit{Inf}}.
\nonumber
\end{eqnarray}
Similarly, the symptomatic case-ICU admission and symptomatic
case-fatality risks are defined as
%
\begin{eqnarray}\label{eqn_sCSR}
\mathit{sCHR}_{w,a} &=& c_{w,a,H | S} ,
\nonumber
\\
\mathit{sCIR}_{w,a} &=& c_{w,a,I | S} =  c_{w,a, I | H} \times
c_{w,a, H | S},
\\
\mathit{sCFR}_{w,a} &=& c_{w,a,D | S}  =  c_{w,a, D | H} \times
c_{w,a, H | S}.
\nonumber
\end{eqnarray}
The conditional probability $c_{w,a,\mathit{Inf} | \mathit{Pop}}$ is commonly referred to
as the ``infection attack rate'' ($\mathit{IAR}_{w,a}$) and $ c_{w,a, S | \mathit{Pop}} =
c_{w,a,S|\mathit{Inf}} \times c_{w,a,\mathit{Inf},\mathit{Pop}}$ is known as the ``symptomatic
attack rate,'' $\mathit{SAR}_{w,a}$.

Let $d_{w,a,l}$ denote ``detection'' probabilities, that is,
probabilities that infections at severity level $l$ are observed. The
full set of wave- and age-specific basic parameters to which we assign
a prior distribution is then
\begin{eqnarray*}
&&\bolds{\theta}(w,a) = \{\mathit{IAR}_{w,a}, c_{w,a, S | \mathit{Inf}},
c_{w,a, H |
S}, c_{w,a, I | H}, c_{w,a, D | H}, d_{w,a,S},
d_{w,a,H},\\
&&\hspace*{262pt} d_{w,a,I}, d_{w,a,D}\},
\end{eqnarray*}
with the total set defined as
\[
\bolds{\Theta} = \bigcup_{w,a} \bolds{\theta}(w,a).
\]

The full set of wave- and age-specific functional parameters is
\begin{eqnarray*}
\bolds{\psi}(w,a) & = & \{ \mathit{SAR}_{w,a},\mathit{CHR}_{w,a},\mathit{sCHR}_{w,a},
\mathit{CIR}_{w,a},\mathit{sCIR}_{w,a},\mathit{CFR}_{w,a},\mathit{sCFR}_{w,a},
\\
& &\hspace*{111pt} N_{w,a,\mathit{Inf}},N_{w,a,S},N_{w,a,H},N_{w,a,I},N_{w,a,D}
\},
\end{eqnarray*}
with the total set defined as
\[
\bolds{\Psi} = \bigcup_{w,a} \bolds{\psi}(w,a).
\]

\subsection{Prior distribution}
The prior distributions assigned to the basic parameters, whether
diffuse or informative, will depend on the specifics of the severity
model considered; see Section~\ref{sec_eng}.

\subsection{Data and likelihood}
In general, at each severity level $l$, we observe $O_{w,a,l}$
infections out of the $N_{w,a,l}$ total infections. Each $O_{w,a,l}$ is
assumed to be Binomially distributed with size parameter $N_{w,a,l}$
and detection probability $d_{w,a,l}$:
\[
O_{w,a,l} \sim \operatorname{Bin}(N_{w,a,l}, d_{w,a,l}).
\]
The likelihood would then be
\[
L(\mathbf{y} | \bolds{\Theta}) = \prod_{w,a,l}
\pmatrix{N_{w,a,l} \cr O_{w,a,l}} d_{w,a,l}^{O_{w,a,l}} (1 -
d_{w,a,l})^{N_{w,a,l} -
O_{w,a,l}}.
\]

The specific models, for example, as in Sections~\ref{sec_model12} and
\ref{sec_3wsep}, may have variations on this likelihood, depending on
the data available. For example, data may be directly available on the
number of hospitalisations resulting in ICU admission, in which case
these data may contribute to the likelihood in the following form:
\[
O_{w,a,I} \sim \operatorname{Bin}(O_{w,a,H}, c_{w,a,I|H}).
\]

\subsection{Computation}
Once the priors and likelihood are defined, samples are obtained from
the resulting joint posterior distribution by MCMC simulation, using
OpenBUGS [\citet{LunnEtAl2009}]. In each model described below, three
independent chains were run for 2,000,000 iterations each, with the
first 500,000 iterations discarded as a burn-in period and the
remainder thinned to every $10$th iteration, resulting in
450,000 samples on which to base posterior inference. Convergence was
established by both visual inspection of the trace plots and
examination of the Brooks--Gelman--Rubin diagnostic plots [\citet{BrooksGelman1998}].

\begin{figure}

\includegraphics{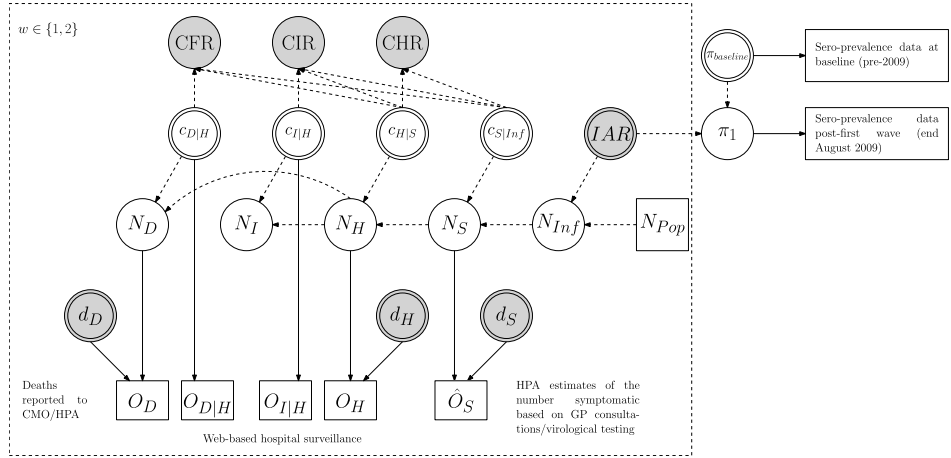}

\caption{Schematic DAG for the severity model, first two waves,
for one age group.}\label{fig_sev12}
\end{figure}

\section{The severity model in England} \label{sec_eng}

The model used in \citet{PresanisEtAl2011a} for the first two waves of
infection in England is described in the next section. Two alternative
methods of modelling the third wave of infection are then given: (a)~a~two-stage approach where posterior distributions from the second wave
model are used to inform prior distributions for some of the
conditional probabilities in the third wave; and (b) a one-stage
approach where all three waves are modelled simultaneously, with the
third wave conditional probabilities parameterised in terms of the
corresponding second wave probabilities.

\subsection{First \& second waves} \label{sec_model12}

Figure~\ref{fig_sev12} is a schematic Directed Acyclic Graph (DAG)
displaying the relationship between parameters and data in the model
for severity in the first two waves in England [\citet{PresanisEtAl2011a}]. The figure displays one generic example age
group, with the $a$ and $w$ indices left out for simplicity. Parameters
are denoted by circles and data by rectangles. The dashed rectangle
represents repetition over the two waves $w \in\{1,2\}$. Double
circles are basic parameters which are assigned prior distributions,
either vague or informative, and filled light grey circles denote the
key parameters (both basic and functional) we wish to estimate. Dashed
arrows denote functional relationships, for example, the definition of
each number $N_{w,a,l} = c_{w,a,l | \lambda} \times N_{w,a,\lambda}$ or
equations (\ref{eqn_CSR}) and (\ref{eqn_sCSR}). Solid arrows represent
distributional assumptions, for example, that an observation is
Binomially distributed.

\subsubsection{Prior distribution}
Independently for each age group, a vague\break $\operatorname{Dirichlet}(1,1,1)$ prior
distribution is given to the infection attack rate,\break $\mathit{IAR}_{w,a}$, in
each of the two waves, together with the remaining fraction of the
population, comprising those either uninfected in the first two waves
or with some degree of immunity at baseline:
\[
\Biggl(\mathit{IAR}_{1,a}, \mathit{IAR}_{2,a}, 1 - \sum
_{w=1}^2 \mathit{IAR}_{w,a} \Biggr) \sim
\operatorname{Dirich}(1,1,1).
\]
The three proportions are therefore constrained {a priori} to sum
to 1 and to lie between 0 and 1. This parameterisation assumes each
infected individual was infected in only a single wave. The remaining
priors are either Uniform or Beta distributions, with full details
given in Section~2.2 of the supplementary material [\citet{PresanisEtAlSupp2013}].

\subsubsection{Likelihood}\label{sec_lik12}
The likelihood is a product of binomial and log-normal contributions,
as detailed in the following.

\paragraph*{Infections}
The sero-prevalence data [source (iii) of Section~\ref{sec_data}] consist of the number of samples testing positive for
pandemic A/H1N1 antibodies, both before and after the first wave. They
are realisations of two binomial distributions and provide information
on the corresponding prevalences at the two time points.
The difference in these two prevalences informs the infection attack
rate in the first wave, via the functional relationship $\pi_{1,a} = \pi
_{\mathrm{baseline},a} + \mathit{IAR}_{1,a}$ (Figure~\ref{fig_sev12}). The post-second
wave sero-prevalence data were not used initially, as some samples
taken after the vaccination campaign had begun were likely to test
positive due to vaccination rather than infection. A lack of
information on the vaccination status of individuals in the sample,
together with concerns that individuals in the sample may have been
more likely than the general population to be at risk of infection, due
to pre-existing conditions, and therefore to be vaccinated [\citet{MillerEtAl2010,Bird2010}], precluded the use of the data without
further work to address these challenges.

\paragraph*{Symptomatic infections}
The estimates $\hat{O}_{w,a,S}$ (Figure~\ref{fig_sev12}) of the number
symptomatic from the HPA (source {(ii)}, Section~1.1.2 of the
supplementary material [\citet{PresanisEtAlSupp2013}]) are assumed to be
log-normally distributed, with a mean that (on the original scale) is
drawn from a binomial distribution with size parameter $N_{w,a,S}$ and
probability parameter given by the detection probability $d_{w,a,S}$.
This parameterisation reflects the belief that the HPA estimates are
underestimates of the number symptomatic $N_{w,a,S}$.

\paragraph*{Hospitalisations and deaths}
The observed hospitalisations $O_{w,a,H}$ and\break deaths $O_{w,a,D}$
[sources {(iv)} and {(v)}, resp., see also Figure~\ref{fig_sev12}] are binomial realisations, with size parameters
$N_{w,a,l}, l \in\{H,D\}$ and probability parameters given by their
respective (wave- but not age-specific) detection probabilities
$d_{w,\ell}$. Amongst observed hospitalisations for whom we have
information on final outcomes [a subset of source {(iv)}], the
observed ICU admissions and deaths are realisations of binomial
distributions with probability parameters given by the conditional
probabilities $c_{w,a,I|H}$ and $c_{w,a,D|H}$, respectively (Figure~\ref{fig_sev12}). Fuller details of the model are given in Section~2 of the
supplementary material [\citet{PresanisEtAlSupp2013}].

\subsection{The third wave: A two-stage approach} \label{sec_3wsep}

\begin{figure}

\includegraphics{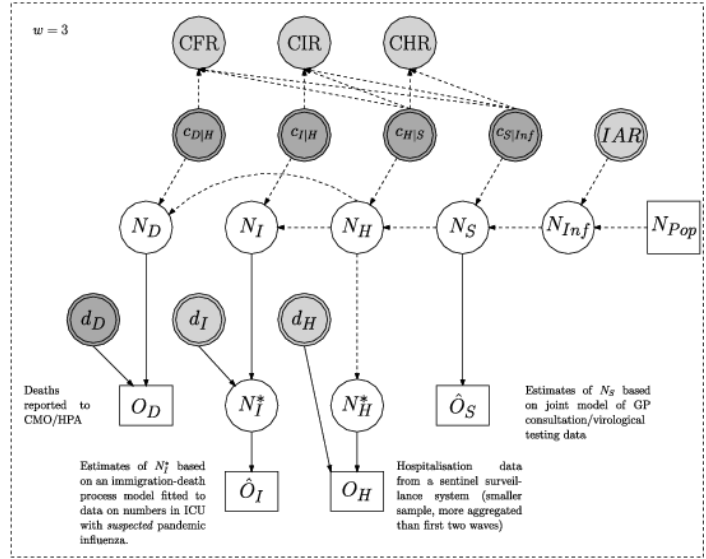}

\caption{Schematic DAG for the severity model, third wave, for one age
group.}\label{fig_sev3}
\end{figure}

The changes in surveillance sources available during the third wave,
particularly the smaller sample sizes and coarser age aggregation,
resulted in the data providing less direct information on the
parameters than in the first two waves. To ensure identifiability of
all parameters, informative prior distributions were employed for some
parameters. The darker grey circles in Figure~\ref{fig_sev3}, a DAG of
the third wave model, denote these parameters, with Beta prior
distributions chosen to reflect the posterior distributions of the
equivalent second wave parameters (see Table~15 of the supplementary
material [\citet{PresanisEtAlSupp2013}]). The changes also entailed two
smaller submodels, one for the data on ICU patients with suspected
pandemic A/H1N1 infection and one for general practice (GP)
consultation and positivity data, the results of which are incorporated
into the third wave severity model as likelihood terms (see below for
more detail).
%

\textit{Infections}.
The infection attack rate again has a Dirichlet prior over the three
waves, but it is now more informative:
\[
(\pi_{2,a},\mathit{IAR}_{3,a},1 - \pi_{2,a} -
\mathit{IAR}_{3,a}) \sim \operatorname{Dirich}(2x_a / y_a, 1, 1),
\]
where $\pi_{2,a}$ is the proportion either with antibodies at baseline
or infected during one of the first two waves, that is, the post-second
wave antibody prevalence. For each age group $a$, $x_a$ and $y_a$ are
chosen such that a $\operatorname{Beta}(x_a,\break y_a)$ distribution approximates the
marginal posterior distribution of $\pi_{2,a}$ derived from the model
of Section~\ref{sec_model12}. The choice of Dirichlet parameters allows
the prior mean for $\pi_{2,a}$ to reflect the posterior mean from
Section~\ref{sec_model12}, but gives greater prior uncertainty than the
corresponding posterior.

\textit{Symptomatic infections.}
As the HPA did not produce estimates of the number symptomatic during
the third wave,
data on ILI consultations and virological positivity from an
alternative primary care sentinel surveillance system ([\citet
{Fleming1999}]; see Section~1.2.1 of the supplementary material
[\citet{PresanisEtAlSupp2013}]) were used to estimate the number symptomatic,
before incorporating this estimate into the severity model. A
log-linear regression of the ILI consultation data on time and age was
fitted jointly with a logistic regression of the positivity data on
time and age [cf. \citet{BirrellEtAl2011}]. A negative binomial
likelihood was assumed for the consultation data and a binomial
likelihood for the positivity data. The number symptomatic due to the
pandemic A/H1N1 strain was then estimated as the sum over weeks of the
product of the expected consultation rate and the expected proportion
positive for pandemic A/H1N1, adjusted for the proportion of
symptomatic patients who contact primary care. The resulting posterior
mean ($\hat{O}^{\prime}_{3,a,S}$) and standard deviation ($\sigma
^{\prime}_{3,a,S}$) of the logarithm of the number symptomatic are
incorporated into the likelihood of the third wave severity model as a
normal term:
\[
\hat{O}^{\prime}_{3,a,S} \sim N \bigl(\log (N_{3,a,S} ),
\sigma ^{\prime 2}_{3,a,S} \bigr)
\]
(see Section~3.1 of the supplementary material [\citet{PresanisEtAlSupp2013}] for details).

\textit{Hospitalisations.}
The hospitalisation data for the third wave (source {(vi)},
Section~1.2.2 of the supplementary material [\citet{PresanisEtAlSupp2013}]) come from a sentinel system. The observed
number of hospitalisations therefore provides a lower bound for the
number of hospitalisations, contributing to the total likelihood as a
binomial component with probability parameter given by the
(non-age specific) detection probability $d_{3,H}$. Recall that these
data are available at a coarser age aggregation than in the first two
waves. The size parameter is therefore a functional parameter
$N^*_{3,b,H} = \sum_{a \in A_{b,H}} N_{3,a,H}$ that is a sum over the
appropriate age groups $a \in A_{b,H}$, where $A_{b,H}$ are sets
describing the mapping from the coarser age groups $b$ to the severity
model age groups $a$.

\textit{ICU admissions.}
The extra information on suspected patients present in ICU (source
{(vii)}, Section~1.2.3 of the supplementary material [\citet{PresanisEtAlSupp2013}]) are modelled as a bivariate immigration-death
process to represent movement in and out of ICU. This process is
combined with the positivity data of source {(viii)} to estimate
the cumulative number of confirmed pandemic A/H1N1 incident cases
admitted to ICU during the third wave (Section~4 of the supplementary
material [\citet{PresanisEtAlSupp2013}]). The resulting posterior mean
(standard deviation) of the logarithm of the cumulative ICU admissions,
$\hat{O}^{\prime}_{3,b,I} (\sigma^{\prime}_{3,b,I})$, are incorporated in the
likelihood for the third wave severity model as normally distributed:
\[
\hat{O}^{\prime}_{3,b,I} \sim N \bigl(\log \bigl(N^*_{3,b,I}
\bigr), \sigma ^{\prime 2}_{3,b,I} \bigr),
\]
where $b$ denotes the age groups available for the suspected ICU data
(two groups: children and adults). As with the hospitalisation data,
the $N^*_{3,b,I} = \sum_{a \in A_{b,I}} N^*_{3,a,I}$ are sums over the
appropriate age groups.
The number $N^*_{w,a,I}$ is still a lower bound for the cumulative
number of ICU admissions over the third wave, since the data of source
{(vii)} cover only a portion of the time of the third wave: this
is expressed as having a binomial distribution with size parameter
$N_{w,a,I}$ and probability parameter given by the age-constant
detection probability $d_{w,I}$.

\textit{Deaths.}
Finally, the observed deaths are again binomially distributed, as in
the first two waves. Full details of the changes to model the third
wave are given in Section~3 of the supplementary material [\citet{PresanisEtAlSupp2013}].

\subsection{Modelling all three waves simultaneously} \label{sec_3wsim}

Modelling the three waves of infection in two stages enables the use of
the posterior distributions of case-severity in the second wave as
prior distributions in the third wave analysis. However, a~two-stage
approach does not allow estimation of the posterior probability of a
change in severity occurring over waves. To do so requires modelling
all three waves simultaneously, as if we had not seen any of the data
until the end of the third wave.

A joint model for all three waves implies different assumptions from
the two-stage approach. First, the prior distribution for the infection
attack rates in each wave is assumed again to be diffuse:
\[
\Biggl(\mathit{IAR}_{1,a},\mathit{IAR}_{2,a},\mathit{IAR}_{3,a},1 - \sum
_{w=1}^3 \mathit{IAR}_{w,a} \Biggr) \sim
\operatorname{Dirich}(1,1,1,1).
\]
Here, the remaining fraction of the population $1 - \sum_{w=1}^3
\mathit{IAR}_{w,a}$ comprises both those with antibodies at baseline
(pre-pandemic) and those remaining uninfected by the end of the third wave.

The proportion symptomatic, $c_{S|\mathit{Inf}}$, is now constrained to be equal
across all three waves and all age groups, instead of its third wave
prior being informed by its second wave posterior distribution.
Likewise, the three conditional probabilities $\mathit{sCHR}_{w,a} =
c_{w,a,H|S}, c_{w,a,I|H}$ and $c_{w,a,D|H}$ for $w = 3$ are no longer
given prior distributions based on second wave posterior distributions,
but are parameterised in terms of their corresponding second wave
conditional probabilities:
%
\begin{eqnarray}\label{eqn_3wParam}
\operatorname{logit}(c_{3,a,l|\lambda}) & \sim& N\bigl(\operatorname {logit}
(c_{2,a,l|\lambda}),
\tau^2_{l|\lambda}\bigr)\nonumber
\\
& & \mbox{for each $(l|\lambda) \in\bigl\{(H|S),(I|H),(D|H)\bigr\}$},
\\
\tau_{l|\lambda} & \sim& \operatorname{Unif}[0,1].
\nonumber
\end{eqnarray}
A value of $\tau= 1$ for the standard deviations would imply that the
odds ratios of the third compared to the second wave probabilities lie
between 0.14 and 7.10. A~value of $\tau= 0$ would imply an odds ratio
of 1, that is, equality of the conditional probabilities:
$c_{3,a,l|\lambda} = c_{2,a,l|\lambda}$.

All other aspects of the joint model for all three waves are as in the
separate first/second and third wave models of Sections~\ref{sec_model12} and \ref{sec_3wsep}, respectively.

\begin{figure}

\includegraphics{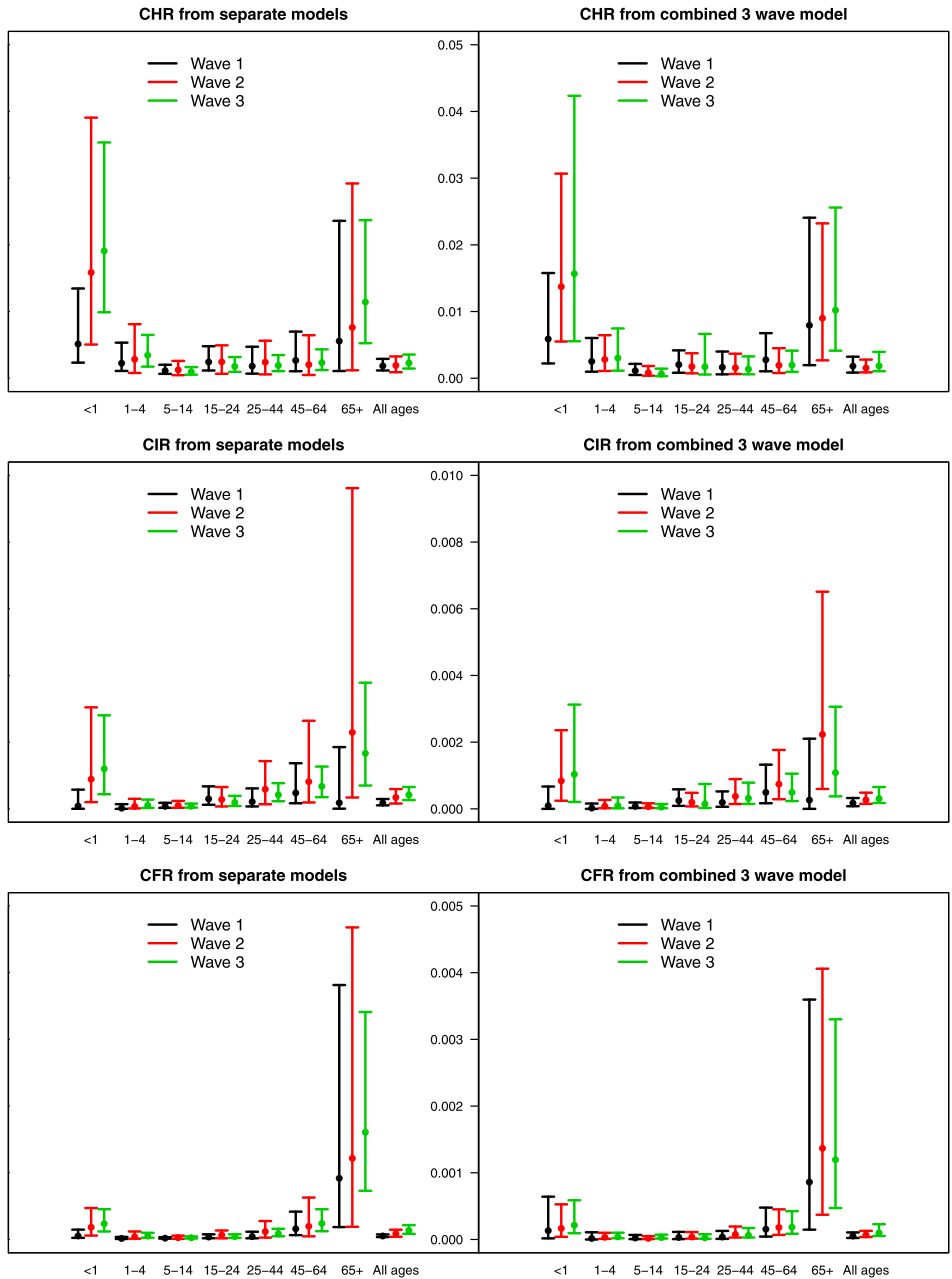}

\caption{$\mathit{CHR},\mathit{CIR}$ and $\mathit{CFR}$ by age, wave and model. Note the different
scales on the $y$-axes.}\label{fig_csrs}
\end{figure}

\begin{figure}

\includegraphics{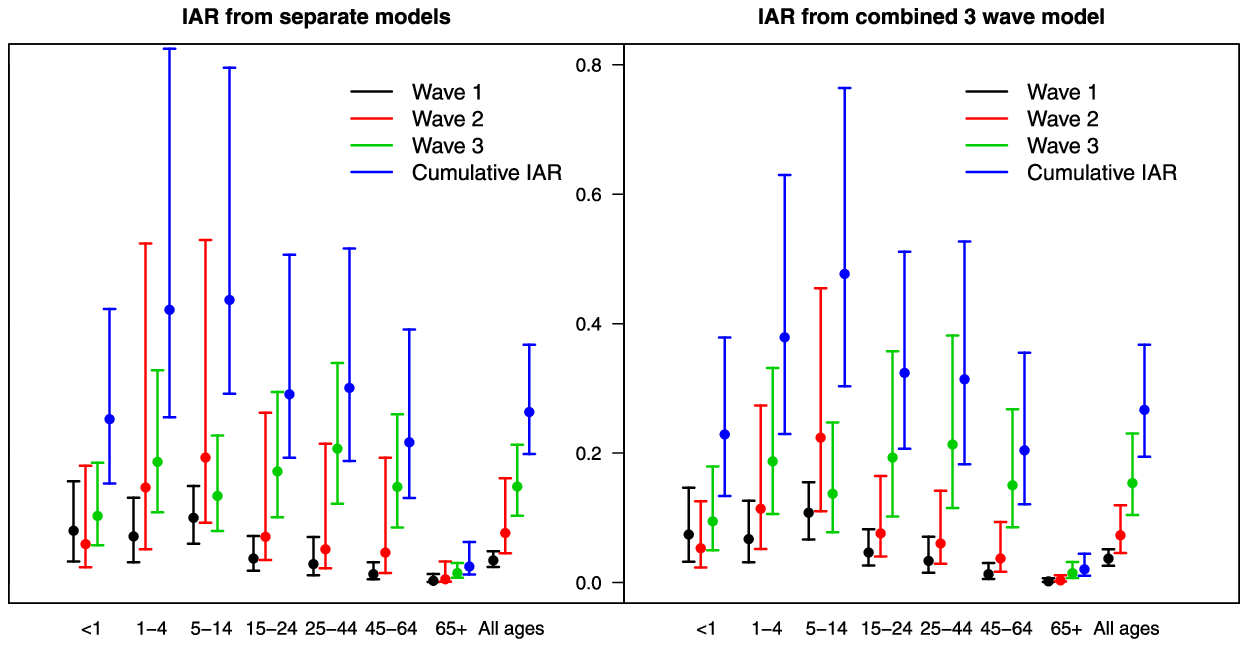}

\caption{Infection attack rate by age, wave and model.}\label{fig_IAR}
\end{figure}

\section{Results} \label{sec_results}

Results from the model for the first two waves, given in full in \citet
{PresanisEtAl2011a}, suggest a mild pandemic, characterised by
case-severity risks increasing between the two waves. From the analysis
of data from the third wave, Figures~\ref{fig_csrs} and \ref{fig_IAR}
show the posterior medians and $95\%$ credible intervals for the
case-severity risks and infection attack rates, respectively, by age,
wave and model. Although there are some differences between the
two-stage models (left-hand sides of the figures) and the combined
three-wave model (right-hand sides), the conclusions are broadly
similar. There is a clear ``u''-shape to the age distribution of the
case-severity risks (Figure~\ref{fig_csrs}) in all three waves, with
the youngest and oldest age groups having the highest probabilities of
experiencing severe events, but also the most uncertainty in the
estimates. The age distribution of the infection attack rates (Figure~\ref{fig_IAR}), on the other hand, is convex, with school-age children
having the highest probability of being infected in the first two
waves, though not the third.

The joint three-wave model allows estimation of the posterior
probabilities of increases across waves in either the attack rates or
the case-severity risks (Table~\ref{tab_ppinc}). Across waves, there is
some evidence of a shift in the age distribution of the infection
attack rates, with posterior probabilities $p \geq0.85$ of an increase
from the second to third waves seen in adults and the very young, but
not in school-age children (posterior probability $p = 0.13$).

At first glance, the estimates averaged over the age groups (Table~\ref{tab_ppinc}) suggest the case-severity risks have increased over all
three waves. The posterior probabilities of a rise across waves of the
$\mathit{CIR}$ and $\mathit{CFR}$ are all greater than $73\%$. However, closer scrutiny
of the age-specific estimates shows this increase does not occur
consistently in every age group and wave. There is stronger evidence of
a rise in ICU admission and fatalities from the first to second waves
than from the second to third. The pattern is less clear in the
case-hospitalisation risks.

\begin{table}
\caption{Posterior probabilities, by age, that the $\mathit{IAR}, \mathit{CHR}, \mathit{CIR}, \mathit{CFR},
c_{I|H}$ and $c_{D|H}$ are greater in \textup{(a)}~wave 2 vs wave 1,
\textup{(b)} wave 3
vs wave 2, and \textup{(c)} wave 3 vs wave 1} \label{tab_ppinc}
\begin{tabular*}{\textwidth}{@{\extracolsep{\fill}}lcd{4.0}d{4.0}d{4.0}@{}}
\hline
& \textbf{Age} & \multicolumn{1}{c}{$\bolds{\operatorname{Pr}(W2 > W1)}$} &
\multicolumn{1}{c}{$\bolds{\operatorname{Pr}(W3 > W2)}$} &
\multicolumn{1}{c@{}}{$\bolds{\operatorname{Pr}(W3 > W1)}$}\\
\hline
{$\mathit{IAR}$} &
$<$1 & $19\%$ & $87\%$ & $71\%$ \\
& 1--4 & $96\%$ & $85\%$ & $99\%$ \\
& 5--14 & $99\%$ & $13\%$ & $77\%$ \\
& 15--24 & $94\%$ & $96\%$ & $100\%$ \\
& 25--44 & $96\%$ & $100\%$ & $100\%$ \\
& 45--64 & $100\%$ & $99\%$ & $100\%$ \\
& 65$+$ & $95\%$ & $99\%$ & $100\%$ \\[3pt]
& All ages & $100\%$ & $99\%$ & $100\%$ \\[6pt]
{$\mathit{CHR}$} &
$<$1 & $ 96\%$ & $62\%$ & $ 95\%$ \\
& 1--4 & $ 60\%$ & $57\%$ & $ 64\%$ \\
& 5--14 & $ 26\%$ & $27\%$ & $ 13\%$ \\
& 15--24 & $ 37\%$ & $48\%$ & $ 39\%$ \\
& 25--44 & $ 47\%$ & $36\%$ & $ 38\%$ \\
& 45--64 & $ 19\%$ & $52\%$ & $ 24\%$ \\
& 65$+$ & $ 60\%$ & $63\%$ & $ 68\%$ \\[3pt]
& All ages &$ 30\%$ & $79\%$ & $54\%$\\[6pt]
{$\mathit{CIR}$} &
$<$1 & $ 98\%$ & $ 62\%$ & $ 97\%$ \\
& 1--4 & $ 86\%$ & $ 58\%$ & $ 86\%$ \\
& 5--14 & $ 43\%$ & $ 34\%$ & $ 31\%$ \\
& 15--24 & $ 34\%$ & $ 31\%$ & $ 27\%$ \\
& 25--44 & $ 91\%$ & $ 36\%$ & $ 82\%$ \\
& 45--64 & $ 80\%$ & $ 14\%$ & $ 50\%$ \\
&  65$+$ & $ 98\%$ & $ 11\%$ & $ 89\%$ \\[3pt]
& All ages &$ 90\%$ & $ 73\%$ & $ 95\%$ \\[6pt]
{$\mathit{CFR}$} &
 $<$1 & $ 58\%$ & $ 69\%$ & $ 70\%$ \\
& 1--4 & $ 74\%$ & $ 64\%$ & $ 80\%$ \\
& 5--14 & $ 38\%$ & $ 82\%$ & $ 66\%$ \\
& 15--24 & $ 60\%$ & $ 24\%$ & $ 41\%$ \\
& 25--44 & $ 86\%$ & $ 35\%$ & $ 77\%$ \\
& 45--64 & $ 61\%$ & $ 52\%$ & $ 61\%$ \\
& 65$+$ & $ 74\%$ & $ 40\%$ & $ 67\%$ \\[3pt]
&       All ages        &$      76\%$   & $     90\%$   & $     93\%$   \\[6pt]
{$c_{I|H}$} &
$<$1 & $ 90\%$ & $ 54\%$ & $ 90\%$ \\
& 1--4 & $ 85\%$ & $ 54\%$ & $ 85\%$ \\
& 5--14 & $ 65\%$ & $ 53\%$ & $ 66\%$ \\
& 15--24 & $ 39\%$ & $ 32\%$ & $ 28\%$ \\
& 25--44 & $ 100\%$ & $ 43\%$ & $ 99\%$ \\
& 45--64 & $ 100\%$ & $ 0\%$ & $ 93\%$ \\
&  65$+$ & $ 99\%$ & $ 1\%$ & $ 87\%$ \\[3pt]
& All ages &$ 100\%$ & $ 35\%$ & $ 100\%$ \\
\hline
\end{tabular*}
\end{table}
\setcounter{table}{0}
\begin{table}
\caption{(Continued)}
\begin{tabular*}{\textwidth}{@{\extracolsep{\fill}}lcd{4.0}d{4.0}d{4.0}@{}}
\hline
& \textbf{Age} & \multicolumn{1}{c}{$\bolds{\operatorname{Pr}(W2 > W1)}$} &
\multicolumn{1}{c}{$\bolds{\operatorname{Pr}(W3 > W2)}$} &
\multicolumn{1}{c@{}}{$\bolds{\operatorname{Pr}(W3 > W1)}$}\\
\hline
{$c_{D|H}$} &
$<$1 & $ 26\%$ & $ 62\%$ & $ 31\%$ \\
&  1--4 & $ 71\%$ & $ 61\%$ & $ 76\%$ \\
&  5--14 & $ 52\%$ & $ 96\%$ & $ 93\%$ \\
& 15--24 & $ 69\%$ & $ 20\%$ & $ 46\%$ \\
& 25--44 & $ 94\%$ & $ 43\%$ & $ 91\%$ \\
& 45--64 & $ 90\%$ & $ 51\%$ & $ 89\%$ \\
&  65$+$ & $ 72\%$ & $ 20\%$ & $ 54\%$ \\[3pt]
& All ages &$ 97\%$ & $ 85\%$ & $ 98\%$ \\
\hline
\end{tabular*}
\end{table}

The reason for the pattern of increase in the $\mathit{CIR}$ and $\mathit{CFR}$ over
waves is not immediately apparent without further investigation. Three
possible hypotheses are as follows: (a) that the increase is due to the
age shift in the infection attack rate away from school-age children
toward adults across waves; (b) that the lack of third wave data and
consequent parameterisation of some of the third wave conditional
probabilities in terms of the corresponding second wave probabilities
[equation~(\ref{eqn_3wParam})] results in the attenuated change in
severity from the second to third wave; and/or (c) that unaccounted
differences in the representativeness of the different surveillance
systems used in the third wave compared to the first two may have an
effect on the estimated severity. These possibilities are not mutually
exclusive and the extent to which the estimated severity is reliant on
each is unknown.

\subsection{Sensitivity analyses} \label{sec_sens12}

The potential for unaccounted biases in the sero-prevalence data
(Sections~\ref{sec_data} and \ref{sec_lik12}), as well as the belief
that the HPA case estimates represented underestimates, prompted
several sensitivity analyses to further assess the uncertainty in the
infection attack rates in the first two waves. Sensitivity to the
choice of data informing the denominators (the infection attack rate
$\mathit{IAR}_{w,a}$ or the number of symptomatic infections $N_{w,a,S}$) and to
the prior distribution of $\mathit{IAR}_{w,a}$ was assessed. Specifically, four
models with different data informing $\mathit{IAR}_{w,a}$ and $N_{w,a,S}$ were
considered:
\begin{longlist}[1.]
\item[1.] using the HPA case estimates to inform $N_{w,a,S}$, assuming they
do so unbiasedly in the first two waves (i.e., with $d_{w,a,S} = 1$),
and using no sero-prevalence data;
\item[2.] the model presented here and in \citet{PresanisEtAl2011a},
assuming the HPA case estimates are biased downwards and using only the
baseline and post-first wave sero-prevalence data;
\item[3.] as in model 2, but using all the sero-prevalence data (up to
post-second wave) of Table~5 of Section~1.1.3 of the supplementary
material [\citet{PresanisEtAlSupp2013}], assuming the HPA case estimates
are biased downwards in both waves; and
\item[4.] as in model 3, but assuming the sero-prevalence data are biased
upwards and the HPA case estimates are biased downwards.
\end{longlist}
Analyses using models 1 and 2 were then repeated using three different
prior distributions for the infection attack rate:
\begin{longlist}[a.]
\item[a.] $\operatorname{Dirichlet}(2,2,6)$, allowing the total attack rate over the two
waves to be a priori $0.4$ on average, with $95\%$ prior mass in
the interval (0.1--0.7), and with a $1:1$ ratio between the two waves;
\item[b.] $\operatorname{Dirichlet}(2.67,1.33,6)$, allowing again a prior total attack
rate of 0.4 (0.1--0.7), but with a $2\dvtx 1$ ratio between waves;
\item[c.] $\operatorname{Dirichlet}(1.33,2.67,6)$, allowing
a prior total attack rate of
0.4 (0.1--0.7), with a $1:2$ ratio between waves.
\end{longlist}

The choice of informative priors is motivated by the total attack rates
in prior pandemics, with the prior uncertainty still relatively large.
\citet{JacksonEtAl2010} found susceptible attack rates (i.e.,
proportion of susceptibles infected, as opposed to proportion of the
total population) of between $19$ and $58\%$ in the first wave of the
1968--1969 pandemic, compared to between $15$ and $50\%$ in the second,
which motivates prior (b). This prior may in fact be sceptical for the
2009 pandemic, as instead of a $2:1$ ratio between waves, the HPA case
estimates and the severe data suggest the ratio was at least $1:1$, if
not $1:2$ or greater. However, this ratio may vary by both age and
region, with London in particular experiencing a somewhat different
epidemic to the rest of the country [\citet{BirrellEtAl2011}]. Prior (c)
therefore allows for the converse, with a greater second wave than first.

The sensitivity analyses to the choice of prior distribution of the
infection attack rate in the first two waves suggest the key messages
from \citet{PresanisEtAl2011a} are robust to the choice of prior
distribution. Results were less robust to the choice of denominator
data included in the model. The inclusion of the post-second wave
sero-prevalence data suggested a higher infection attack rate [28.4\%\
(26.0--30.8\%)] than the baseline analysis [11.2\%\ (7.4--18.9\%)],
with a corresponding lower case-fatality risk in the second wave
[0.0027\%\ (0.0024--0.0031\%) compared to 0.009\%\ (0.004--0.014\%)].
Full details of these sensitivity analyses are given in Section~5 of
the supplementary material [\citet{PresanisEtAlSupp2013}]. Recall
(Section~\ref{sec_lik12}) that the samples tested post-second wave and
before and after the third wave [\citet{HoschlerEtAl2012}] may
overrepresent individuals at higher risk of infection and vaccination.
The observed sero-prevalence in these samples may therefore suggest a
higher infection attack rate than truly occurred. Further work to
obtain background information on individuals in the samples, and
therefore to account for sampling biases, is underway, prompted in part
by the results of these sensitivity analyses.

\begin{figure}

\includegraphics{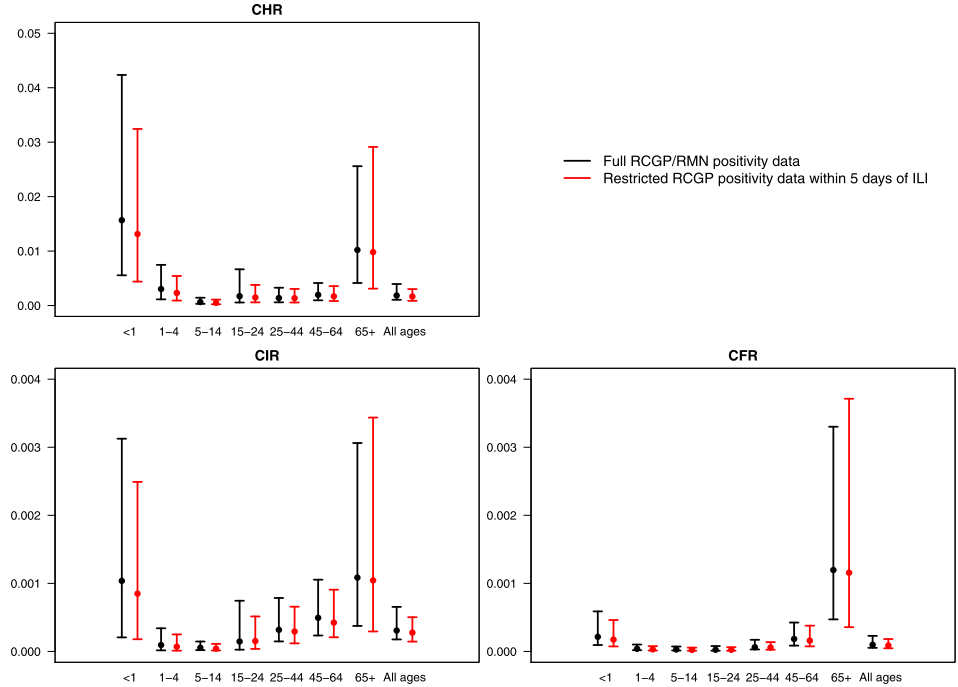}

\caption{Third wave $\mathit{CHR},\mathit{CIR}$ and $\mathit{CFR}$ from the combined model, by age
and source of positivity data. Note the different scales on the $y$-axes.}
\label{fig_csrs_sens}\vspace*{-3pt}
\end{figure}

\begin{figure}

\includegraphics{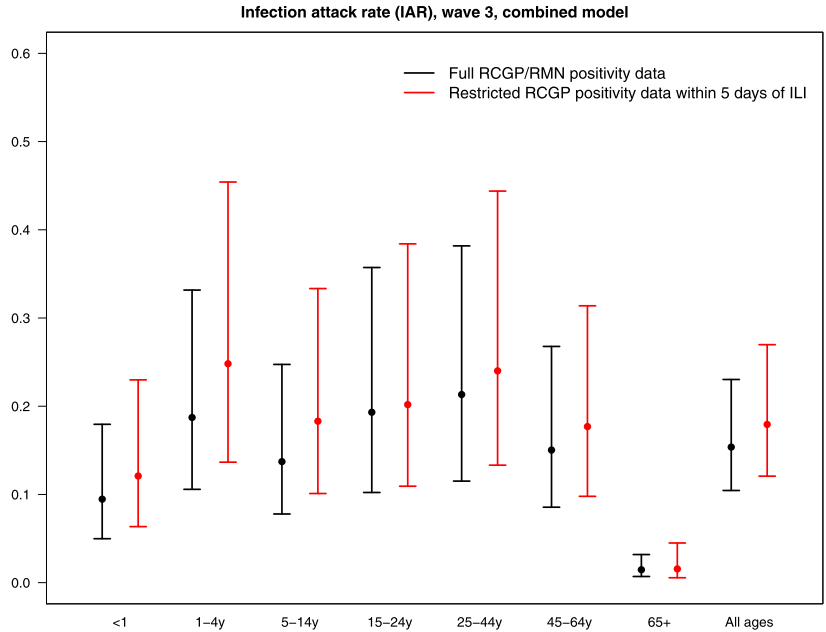}

\caption{Third wave infection attack rate from the combined model, by
age and positivity data set used.}\label{fig_IAR_sens}
\end{figure}

In the third wave, a sensitivity analysis to the set of virological
positivity data used was performed (Sections~1.2.1 and 3.1 of the
supplementary material [\citet{PresanisEtAlSupp2013}]). The main
analysis used the full positivity data, with the results of the
Bayesian joint regression model of the positivity and primary care
consultation data (Table~13 of the supplementary material) incorporated
into the combined 3-wave model as shown in Figure~\ref{fig_sev3} and
Section~3.1 of the supplementary material. The sensitivity analysis
employed instead a set of virological positivity data restricted to
tests made on swabs that were collected within 5 days of an ILI
consultation, with corresponding results from the joint regression
model in Table~14\vadjust{\goodbreak} of the supplementary material. The results from
including the two alternative sets of estimates from the joint
regression model into the combined 3-wave model are compared in Figures~\ref{fig_csrs_sens} and \ref{fig_IAR_sens}.
%

The general conclusions about the age distribution of the case-severity risks and
infection attack rate in the third wave are unchanged by the use of the
restricted positivity data. The restricted data do imply a slightly
higher and more uncertain attack rate in each age group (Figure~\ref{fig_IAR_sens}), due to the higher observed positivity and smaller
sample sizes. Correspondingly, the case-severity risks are slightly
lower in each age group in the sensitivity analysis (Figure~\ref{fig_csrs_sens}), but the greater uncertainty in the denominator does
not seem to translate directly into greater uncertainty in the risks.

\section{Discussion} \label{sec_discuss}

We have extended and further developed a Bayesian evidence synthesis
model [\citet{PresanisEtAl2011a}] to characterise and estimate the
severity of the 2009 pandemic A/H1N1 strain of influenza in the three
waves of infection experienced in England. The model has been adapted
to account for changes in the surveillance data available over the
course of the three waves, considering two approaches: (a) a two-stage
approach, using posterior distributions from the model for the first
two waves to inform prior distributions for the third wave analysis;
and (b) modelling all three waves simultaneously,\vadjust{\goodbreak} accounting for the
reduction in available data by parameterising the third wave severity
parameters in terms of the corresponding second wave parameters. Both
approaches have resulted in broadly the same three key conclusions:
\begin{longlist}[1.]
\item[1.] The age distribution in case-severity risks is ``u''-shaped,
implying children aged less than a year and older adults have highest
severity, although their estimates are also the most uncertain. This
pattern is consistent with the increasing severity with age seen in
other countries during the 2009 pandemic [\citet{PresanisEtAl2009,WuEtAl2010,SypsaEtAl2011,ShubinEtAl2013,WongEtAl2013}],
where in each of these analyses, the authors did not distinguish
between children under 1 year of age and those aged 1--4. The pattern is
also consistent with global relative risks by age of severe events
compared to the general population estimated by \citet{VanKerkhoveEtAl2011}.
\item[2.] The age distribution of the infection attack rate changes over
waves, with school-age children most affected in the first two waves
and an increase in the attack rate in adults aged 25 and older from the
second to third waves.
\item[3.] When averaged over all ages, severity in those infected appears
to increase over the three waves.
\end{longlist}
The changing age distribution and apparent increase in severity over
waves is consistent with estimates from the two pandemic waves
experienced by other countries [\citet{YangEtAl2011,TrueloveEtAl2011,ShubinEtAl2013}].

It is important to note that the estimates presented here do not
account for risk factors for severe influenza, nor for vaccination
status nor for other preventive measures, such as social distancing,
which might have an effect on severity. Both the joint regression model
of virological positivity and GP consultation and the full severity
model would require further development to account for these factors
and to be able to use the second and third wave serology data
accounting for sampling biases. Assessment of the effect on estimates
of assumptions---such as that of no influenza-related deaths occurring
outside of hospital or the parameterisation of the third wave in terms
of the second wave in the combined analysis---is also key. The
possible effect of any differences in representativeness of the various
surveillance systems in the third compared to the first two waves is an
issue for further investigation. The sample sizes and prior
distributions chosen do not provide enough information to enable
convergence of the MCMC algorithm for the severity model when taking
the number of infections $N_{w,a,l}$ to be a Binomial realisation from
the number at a less severe level $\lambda$. This lack of convergence
implies there may be too much uncertainty to allow identifiability of
the model in this case, prompting instead the mean assumption
$N_{w,a,l} = c_{w,a,l | \lambda} \times N_{w,a,\lambda}$. Another area
for future investigation is to assess how informative the priors are
required to be or how large sample sizes need to be to enable
convergence when the $N_{w,a,\lambda}$ are stochastic.

Despite these challenges, our Bayesian evidence synthesis approach has
allowed us to draw important public health conclusions, not only in
characterising the severity of the 2009 pandemic, but also in shaping
future research. The sensitivity analyses showed the severity estimates
were robust to prior assumptions about the infection attack rate, but
less robust to the choice of data to include in informing the attack
rate. Although the magnitude of the severity estimates varied, the
conclusions of a ``u''-shaped age distribution to severity and an
apparent increase in severity over waves were nevertheless robust. The
sensitivity of the results has, furthermore, contributed to the
initiation of a project to obtain further data to better understand the
potential sampling bias in the sero-prevalence data [\citet{LaurieEtAl2012}].

The evidence synthesis framework has also given us the flexibility to
account for biases, using prior information on parameters representing
the biases, for example. Bias modelling has been an integral part of
the model development, inference and criticism cycle, as have the
sensitivity analyses. It is important, in any analysis, to understand
the contribution of each item of evidence, whether in the form of model
structure, prior distribution or data, in driving inferences. It is
particularly crucial when an analysis relies on informative priors for
identifiability, as is the case here. Another key aspect of model
criticism in an evidence synthesis is to assess the consistency of the
various data sources, not only with each other, but also with the model
structure. It is possible, and indeed common, in syntheses of multiple
sources of evidence to find both that some parameters are only barely
identified by the data and that other parameters are informed
indirectly by more than one data item. In the latter case, there is
clearly potential for different sources of data to conflict, providing
inconsistent evidence on a particular parameter [\citet{LuAdes2006,SweetingEtAl2008,PresanisEtAl2008}]. Such conflicts need to
be detected, measured, understood and resolved. Conflict diagnostics,
in the form of cross-validatory posterior prediction, for the first
wave confirm the inconsistency between the serology data and the HPA
estimates of the number symptomatic if taken at face value [\citet{PresanisEtAl2013}]. In our main analysis, we addressed the conflict by
incorporating a bias parameter for the HPA estimates, whereas in the
sensitivity analyses, we also considered a bias parameter for the
serology data. Further preliminary work on measuring conflict seems to
confirm the suggestion of the sensitivity analyses that the severe end
data does indeed conflict with the evidence on the attack rates. Given
the uncertainties in the attack rates, understanding and resolving this
conflict is an important next step. The iterative process of fitting,
criticising and further developing an evidence synthesis model to
address conflicts, as we have done and are continuing to do here, leads
automatically to internal consistency. By contrast, external validation
is much more challenging in an evidence synthesis framework. As already
noted, due to identifiability issues common to evidence syntheses, it
is rare to find external data against which to validate---such data are
instead used in the synthesis.

Despite these challenges, an evidence synthesis using a complex
probabilistic model provides a powerful approach to estimating
influenza severity when the available evidence comes from multiple
sources that are incomplete and biased. The embedding of a ``pyramid''
approach to severity estimation within an evidence synthesis framework,
as presented here, is easily adapted to other contexts, both within
epidemiology, where many diseases may be observed at different levels
of severity or diagnosis, and in other fields where observation occurs
at different levels, for example, quality control or ecology.

\section*{Acknowledgements}
We thank colleagues at Public Health England and the Royal College of
General Practitioners, particularly Michele Barley, who provided data
for this analysis. We are grateful to all GPs who participate in the
RCGP Weekly Returns Service. We would particularly like to acknowledge
Professor J.~R. Norris (University of Cambridge) and Professor Marc
Lipsitch (Harvard School of Public Health) for advice on early versions
of this analysis.


\begin{supplement}[id=suppA]
\stitle{Appendix: Synthesising evidence to estimate pandemic (2009)
A/H1N1 influenza severity in 2009--2011}
\slink[doi]{10.1214/14-AOAS775SUPP} 
\sdatatype{.pdf}
\sfilename{aoas775\_supp.pdf}
\sdescription{Appendix describing the data, further model details
and sensitivity analyses.}
\end{supplement}





\printaddresses

\begin{thebibliography}{47}

\bibitem[\protect\citeauthoryear{Ades and Sutton}{2006}]{AdesSutton2006}
\begin{barticle}[mr]
\bauthor{\bsnm{Ades},~\bfnm{A.~E.}\binits{A.~E.}} \AND
\bauthor{\bsnm{Sutton},~\bfnm{A.~J.}\binits{A.~J.}}
(\byear{2006}).
\btitle{Multiparameter evidence synthesis in epidemiology and medical decision-making: Current approaches}.
\bjournal{J. Roy. Statist. Soc. Ser. A}
\bvolume{169}
\bpages{5--35}.
\bid{doi={10.1111/j.1467-985X.2005.00377.x}, issn={0964-1998}, mr={2222010}}
\end{barticle}
%
\bptok{imsref}%
\endbibitem

\bibitem[\protect\citeauthoryear{Albert et~al.}{2011}]{AlbertEtAl2011}
\begin{barticle}[author]
\bauthor{\bsnm{Albert},~\bfnm{Isabelle}\binits{I.}},
\bauthor{\bsnm{Espi{\'{e}}},~\bfnm{Emmanuelle}\binits{E.}},
\bauthor{\bparticle{de} \bsnm{Valk},~\bfnm{Henriette}\binits{H.}} \AND
\bauthor{\bsnm{Denis},~\bfnm{Jean-Baptiste~B.}\binits{J.-B.~B.}}
(\byear{2011}).
\btitle{{A Bayesian evidence synthesis for estimating campylobacteriosis prevalence.}}
\bjournal{Risk Analysis}
\bvolume{31}
\bpages{1141--1155}.
\end{barticle}
%
\bptok{imsref}%
\endbibitem

\bibitem[\protect\citeauthoryear{Bird}{2010}]{Bird2010}
\begin{barticle}[author]
\bauthor{\bsnm{Bird},~\bfnm{Sheila~M.}\binits{S.~M.}}
(\byear{2010}).
\btitle{Like-with-like comparisons?}
\bjournal{The Lancet}
\bvolume{376}
\bpages{684+}.
\end{barticle}
%
\bptok{imsref}%
\endbibitem

\bibitem[\protect\citeauthoryear{Birrell et~al.}{2011}]{BirrellEtAl2011}
\begin{barticle}[author]
\bauthor{\bsnm{Birrell},~\bfnm{Paul~J.}\binits{P.~J.}},
\bauthor{\bsnm{Ketsetzis},~\bfnm{Georgios}\binits{G.}},
\bauthor{\bsnm{Gay},~\bfnm{Nigel~J.}\binits{N.~J.}},
\bauthor{\bsnm{Cooper},~\bfnm{Ben~S.}\binits{B.~S.}},
\bauthor{\bsnm{Presanis},~\bfnm{Anne~M.}\binits{A.~M.}},
\bauthor{\bsnm{Harris},~\bfnm{Ross~J.}\binits{R.~J.}},
\bauthor{\bsnm{Charlett},~\bfnm{Andr{\'{e}}}\binits{A.}},
\bauthor{\bsnm{Zhang},~\bfnm{Xu-Sheng}\binits{X.-S.}},
\bauthor{\bsnm{White},~\bfnm{Peter~J.}\binits{P.~J.}},
\bauthor{\bsnm{Pebody},~\bfnm{Richard~G.}\binits{R.~G.}} \AND
\bauthor{\bsnm{De~Angelis},~\bfnm{Daniela}\binits{D.}}
(\byear{2011}).
\btitle{Bayesian modeling to unmask and predict influenza {A/H1N1pdm} dynamics in London}.
\bjournal{Proc. Natl. Acad. Sci. USA}
\bvolume{108}
\bpages{18238--18243}.
\end{barticle}
%
\bptok{imsref}%
\endbibitem

\bibitem[\protect\citeauthoryear{Box}{1980}]{Box1980}
\begin{barticle}[mr]
\bauthor{\bsnm{Box},~\bfnm{George~E.~P.}\binits{G.~E.~P.}}
(\byear{1980}).
\btitle{Sampling and {B}ayes' inference in scientific modelling and robustness}.
\bjournal{J. Roy. Statist. Soc. Ser. A}
\bvolume{143}
\bpages{383--430}.
\bid{doi={10.2307/2982063}, issn={0035-9238}, mr={0603745}}
\bptnote{check related}%
\end{barticle}
%
\bptok{imsref}%
\endbibitem

\bibitem[\protect\citeauthoryear{Brooks and Gelman}{1998}]{BrooksGelman1998}
\begin{barticle}[mr]
\bauthor{\bsnm{Brooks},~\bfnm{Stephen~P.}\binits{S.~P.}} \AND
\bauthor{\bsnm{Gelman},~\bfnm{Andrew}\binits{A.}}
(\byear{1998}).
\btitle{General methods for monitoring convergence of iterative simulations}.
\bjournal{J. Comput. Graph. Statist.}
\bvolume{7}
\bpages{434--455}.
\bid{doi={10.2307/1390675}, issn={1061-8600}, mr={1665662}}
\end{barticle}
\bptok{imsref}%
\endbibitem

\bibitem[\protect\citeauthoryear{Campbell et~al.}{2011}]{CampbellEtAl2011}
\begin{barticle}[author]
\bauthor{\bsnm{Campbell},~\bfnm{C.~N.~J.}\binits{C.~N.~J.}},
\bauthor{\bsnm{Mytton},~\bfnm{O.~T.}\binits{O.~T.}},
\bauthor{\bsnm{McLean},~\bfnm{E.~M.}\binits{E.~M.}},
\bauthor{\bsnm{Rutter},~\bfnm{P.~D.}\binits{P.~D.}},
\bauthor{\bsnm{Pebody},~\bfnm{R.~G.}\binits{R.~G.}},
\bauthor{\bsnm{Sachedina},~\bfnm{N.}\binits{N.}},
\bauthor{\bsnm{White},~\bfnm{P.~J.}\binits{P.~J.}},
\bauthor{\bsnm{Hawkins},~\bfnm{C.}\binits{C.}},
\bauthor{\bsnm{Evans},~\bfnm{B.}\binits{B.}},
\bauthor{\bsnm{Waight},~\bfnm{P.~A.}\binits{P.~A.}},
\bauthor{\bsnm{Ellis},~\bfnm{J.}\binits{J.}},
\bauthor{\bsnm{Bermingham},~\bfnm{A.}\binits{A.}},
\bauthor{\bsnm{Donaldson},~\bfnm{L.~J.}\binits{L.~J.}} \AND
\bauthor{\bsnm{Catchpole},~\bfnm{M.}\binits{M.}}
(\byear{2011}).
\btitle{{Hospitalization in two waves of pandemic influenza A(H1N1) in England}}.
\bjournal{Epidemiology and Infection}
\bvolume{139}
\bpages{1560--1569}.
\end{barticle}
%
\bptok{imsref}%
\endbibitem

\bibitem[\protect\citeauthoryear{{Department of Health}}{2011}]{DHWW2011}
\begin{bmisc}[author]
\borganization{Department of Health}
(\byear{2011}).
\bhowpublished{Department of Health Winter Watch.
Accessed 25/02/2011.}
\end{bmisc}
%
\bptok{imsref}%
\endbibitem

\bibitem[\protect\citeauthoryear{Donaldson et~al.}{2009}]{DonaldsonEtAl2009}
\begin{barticle}[pbm]
\bauthor{\bsnm{Donaldson},~\bfnm{Liam~J.}\binits{L.~J.}},
\bauthor{\bsnm{Rutter},~\bfnm{Paul~D.}\binits{P.~D.}},
\bauthor{\bsnm{Ellis},~\bfnm{Benjamin~M.}\binits{B.~M.}},
\bauthor{\bsnm{Greaves},~\bfnm{Felix~E.~C.}\binits{F.~E.~C.}},
\bauthor{\bsnm{Mytton},~\bfnm{Oliver~T.}\binits{O.~T.}},
\bauthor{\bsnm{Pebody},~\bfnm{Richard~G.}\binits{R.~G.}} \AND
\bauthor{\bsnm{Yardley},~\bfnm{Iain~E.}\binits{I.~E.}}
(\byear{2009}).
\btitle{Mortality from pandemic A/H1N1 2009 influenza in England: Public health surveillance study}.
\bjournal{BMJ}
\bvolume{339}
\bpages{b5213}.
\bid{issn={1756-1833}, pmcid={2791802}, pmid={20007665}}
\end{barticle}
%
\bptok{imsref}%
\endbibitem

\bibitem[\protect\citeauthoryear{Eddy, Hasselblad and Shachter}{1992}]{EddyEtAl1992}
\begin{bbook}[author]
\bauthor{\bsnm{Eddy},~\bfnm{David~M.}\binits{D.~M.}},
\bauthor{\bsnm{Hasselblad},~\bfnm{Vic}\binits{V.}} \AND
\bauthor{\bsnm{Shachter},~\bfnm{Ross}\binits{R.}}
(\byear{1992}).
\btitle{{Meta-Analysis by the Confidence Profile Method}}.
\bpublisher{Academic Press},
\blocation{Boston, MA}.
\end{bbook}
%
\bptok{imsref}%
\endbibitem

\bibitem[\protect\citeauthoryear{Fleming}{1999}]{Fleming1999}
\begin{barticle}[author]
\bauthor{\bsnm{Fleming},~\bfnm{D.~M.}\binits{D.~M.}}
(\byear{1999}).
\btitle{{Weekly returns service of the royal college of general practitioners.}}
\bjournal{Communicable Disease and Public Health / PHLS}
\bvolume{2}
\bpages{96--100}.
\end{barticle}
%
\bptok{imsref}%
\endbibitem

\bibitem[\protect\citeauthoryear{Garske et~al.}{2009}]{GarskeEtAl2009}
\begin{barticle}[pbm]
\bauthor{\bsnm{Garske},~\bfnm{Tini}\binits{T.}},
\bauthor{\bsnm{Legrand},~\bfnm{Judith}\binits{J.}},
\bauthor{\bsnm{Donnelly},~\bfnm{Christl~A.}\binits{C.~A.}},
\bauthor{\bsnm{Ward},~\bfnm{Helen}\binits{H.}},
\bauthor{\bsnm{Cauchemez},~\bfnm{Simon}\binits{S.}},
\bauthor{\bsnm{Fraser},~\bfnm{Christophe}\binits{C.}},
\bauthor{\bsnm{Ferguson},~\bfnm{Neil~M.}\binits{N.~M.}} \AND
\bauthor{\bsnm{Ghani},~\bfnm{Azra~C.}\binits{A.~C.}}
(\byear{2009}).
\btitle{Assessing the severity of the novel influenza A/H1N1 pandemic}.
\bjournal{BMJ}
\bvolume{339}
\bpages{b2840}.
\bid{issn={1756-1833}, pmid={19602714}}
\end{barticle}
\bptok{imsref}%
\endbibitem

\bibitem[\protect\citeauthoryear{Goubar et~al.}{2008}]{GoubarEtAl2008}
\begin{barticle}[mr]
\bauthor{\bsnm{Goubar},~\bfnm{A.}\binits{A.}},
\bauthor{\bsnm{Ades},~\bfnm{A.~E.}\binits{A.~E.}},
\bauthor{\bsnm{De Angelis},~\bfnm{D.}\binits{D.}},
\bauthor{\bsnm{McGarrigle},~\bfnm{C.~A.}\binits{C.~A.}},
\bauthor{\bsnm{Mercer},~\bfnm{C.~H.}\binits{C.~H.}},
\bauthor{\bsnm{Tookey},~\bfnm{P.~A.}\binits{P.~A.}},
\bauthor{\bsnm{Fenton},~\bfnm{K.}\binits{K.}} \AND
\bauthor{\bsnm{Gill},~\bfnm{O.~N.}\binits{O.~N.}}
(\byear{2008}).
\btitle{Estimates of human immunodeficiency virus prevalence and proportion diagnosed based on {B}ayesian multiparameter synthesis of surveillance data}.
\bjournal{J. Roy. Statist. Soc. Ser. A}
\bvolume{171}
\bpages{541--580}.
\bid{doi={10.1111/j.1467-985X.2007.00537.x}, issn={0964-1998}, mr={2432503}}
\end{barticle}
%
\bptok{imsref}%
\endbibitem

\bibitem[\protect\citeauthoryear{Hardelid et~al.}{2011}]{HardelidEtAl2011}
\begin{barticle}[author]
\bauthor{\bsnm{Hardelid},~\bfnm{P.}\binits{P.}},
\bauthor{\bsnm{Andrews},~\bfnm{N.~J.}\binits{N.~J.}},
\bauthor{\bsnm{Hoschler},~\bfnm{K.}\binits{K.}},
\bauthor{\bsnm{Stanford},~\bfnm{E.}\binits{E.}},
\bauthor{\bsnm{Baguelin},~\bfnm{M.}\binits{M.}},
\bauthor{\bsnm{Waight},~\bfnm{P.~A.}\binits{P.~A.}},
\bauthor{\bsnm{Zambon},~\bfnm{M.}\binits{M.}} \AND
\bauthor{\bsnm{Miller},~\bfnm{E.}\binits{E.}}
(\byear{2011}).
\btitle{Assessment of baseline age-specific antibody prevalence and incidence of infection to novel influenza {A/H1N1} 2009.}
\bjournal{Health Technology Assessment}
\bvolume{14}
\bpages{115--192}.
\end{barticle}
\bptok{imsref}%
\endbibitem

\bibitem[\protect\citeauthoryear{{Health Protection Agency}}{2010}]{HPAepiReport2010}
\begin{bmisc}[author]
\borganization{Health Protection Agency}
(\byear{2010}).
\bhowpublished{Epidemiological report of pandemic (H1N1) 2009 in the UK.
Technical report, Health Protection Agency.}
\end{bmisc}
%
\bptok{imsref}%
\endbibitem

\bibitem[\protect\citeauthoryear{Health Protection Agency,
Health Protection Scotland,
Communicable Disease Surveillance Centre Northern Ireland and
National Public Health Service for Wales}{2009}]{FF100project2009}
\begin{bmisc}[author]
\borganization{Health Protection Agency,
Health Protection Scotland,
Communicable Disease Surveillance Centre Northern Ireland and
National Public Health Service for Wales}
(\byear{2009}).
\bhowpublished{{First Few Hundred (FF100) Project: Epidemiological
Protocols for Comprehensive Assessment of Early Swine Influenza
Cases in the United Kingdom} {Technical report}, Health Protection Agency.}
\end{bmisc}
\bptok{imsref}%
\endbibitem

\bibitem[\protect\citeauthoryear{Hoschler et~al.}{2012}]{HoschlerEtAl2012}
\begin{barticle}[pbm]
\bauthor{\bsnm{Hoschler},~\bfnm{Katja}\binits{K.}},
\bauthor{\bsnm{Thompson},~\bfnm{Catherine}\binits{C.}},
\bauthor{\bsnm{Andrews},~\bfnm{Nick}\binits{N.}},
\bauthor{\bsnm{Galiano},~\bfnm{Monica}\binits{M.}},
\bauthor{\bsnm{Pebody},~\bfnm{Richard}\binits{R.}},
\bauthor{\bsnm{Ellis},~\bfnm{Joanna}\binits{J.}},
\bauthor{\bsnm{Stanford},~\bfnm{Elaine}\binits{E.}},
\bauthor{\bsnm{Baguelin},~\bfnm{Marc}\binits{M.}},
\bauthor{\bsnm{Miller},~\bfnm{Elizabeth}\binits{E.}} \AND
\bauthor{\bsnm{Zambon},~\bfnm{Maria}\binits{M.}}
(\byear{2012}).
\btitle{Seroprevalence of influenza A(H1N1)pdm09 virus antibody, England, 2010 and 2011}.
\bjournal{Emerging Infect. Dis.}
\bvolume{18}
\bpages{1894--1897}.
\bid{doi={10.3201/eid1811.120720}, issn={1080-6059}, pmcid={3559155}, pmid={23092684}}
\end{barticle}
%
\bptok{imsref}%
\endbibitem

\bibitem[\protect\citeauthoryear{Jackson, Vynnycky and Mangtani}{2010}]{JacksonEtAl2010}
\begin{barticle}[pbm]
\bauthor{\bsnm{Jackson},~\bfnm{Charlotte}\binits{C.}},
\bauthor{\bsnm{Vynnycky},~\bfnm{Emilia}\binits{E.}} \AND
\bauthor{\bsnm{Mangtani},~\bfnm{Punam}\binits{P.}}
(\byear{2010}).
\btitle{Estimates of the transmissibility of the 1968 (Hong Kong) influenza pandemic: Evidence of increased transmissibility between successive waves}.
\bjournal{Am. J. Epidemiol.}
\bvolume{171}
\bpages{465--478}.
\bid{doi={10.1093/aje/kwp394}, issn={1476-6256}, pii={kwp394}, pmcid={2816729}, pmid={20007674}}
\end{barticle}
\bptok{imsref}%
\endbibitem

\bibitem[\protect\citeauthoryear{Laurie et~al.}{2013}]{LaurieEtAl2012}
\begin{barticle}[pbm]
\bauthor{\bsnm{Laurie},~\bfnm{Karen~L.}\binits{K.~L.}},
\bauthor{\bsnm{Huston},~\bfnm{Patricia}\binits{P.}},
\bauthor{\bsnm{Riley},~\bfnm{Steven}\binits{S.}},
\bauthor{\bsnm{Katz},~\bfnm{Jacqueline~M.}\binits{J.~M.}},
\bauthor{\bsnm{Willison},~\bfnm{Donald~J.}\binits{D.~J.}},
\bauthor{\bsnm{Tam},~\bfnm{John~S.}\binits{J.~S.}},
\bauthor{\bsnm{Mounts},~\bfnm{Anthony~W.}\binits{A.~W.}},
\bauthor{\bsnm{Hoschler},~\bfnm{Katja}\binits{K.}},
\bauthor{\bsnm{Miller},~\bfnm{Elizabeth}\binits{E.}},
\bauthor{\bsnm{Vandemaele},~\bfnm{Kaat}\binits{K.}},
\bauthor{\bsnm{Broberg},~\bfnm{Eeva}\binits{E.}},
\bauthor{\bsnm{Van Kerkhove},~\bfnm{Maria~D.}\binits{M.~D.}} \AND
\bauthor{\bsnm{Nicoll},~\bfnm{Angus}\binits{A.}}
(\byear{2013}).
\btitle{Influenza serological studies to inform public health action: Best practices to optimise timing, quality and reporting}.
\bjournal{Influenza Other Respir Viruses}
\bvolume{7}
\bpages{211--224}.
\bid{doi={10.1111/j.1750-2659.2012.0370a.x}, issn={1750-2659}, pmid={22548725}}
\end{barticle}
\bptok{imsref}%
\endbibitem

\bibitem[\protect\citeauthoryear{Lipsitch et~al.}{2009}]{LipsitchEtAl2009}
\begin{barticle}[author]
\bauthor{\bsnm{Lipsitch},~\bfnm{Marc}\binits{M.}},
\bauthor{\bsnm{Riley},~\bfnm{Steven}\binits{S.}},
\bauthor{\bsnm{Cauchemez},~\bfnm{Simon}\binits{S.}},
\bauthor{\bsnm{Ghani},~\bfnm{Azra~C.}\binits{A.~C.}} \AND
\bauthor{\bsnm{Ferguson},~\bfnm{Neil~M.}\binits{N.~M.}}
(\byear{2009}).
\btitle{{Managing and reducing uncertainty in an emerging influenza pandemic}}.
\bjournal{New England Journal of Medicine}
\bvolume{361}
\bpages{112--115}.
\end{barticle}
\bptok{imsref}%
\endbibitem

\bibitem[\protect\citeauthoryear{Lipsitch et~al.}{2011}]{LipsitchEtAl2011}
\begin{barticle}[author]
\bauthor{\bsnm{Lipsitch},~\bfnm{M.}\binits{M.}},
\bauthor{\bsnm{Finelli},~\bfnm{L.}\binits{L.}},
\bauthor{\bsnm{Heffernan},~\bfnm{R.~T.}\binits{R.~T.}},
\bauthor{\bsnm{Leung},~\bfnm{G.~M.}\binits{G.~M.}},
\bauthor{\bsnm{Redd},~\bfnm{S.~C.}\binits{S.~C.}} \AND
\bauthor{\bsnm{{for the 2009 H1N1 Surveillance Group}}}
(\byear{2011}).
\btitle{Improving the evidence base for decision making during a pandemic: The example of 2009 influenza {A/H1N1}}.
\bjournal{Biosecurity and Bioterrorism: Biodefense Strategy, Practice, and Science}
\bvolume{9}
\bpages{89--114}.
\end{barticle}
%
\bptok{imsref}%
\endbibitem

\bibitem[\protect\citeauthoryear{Lu and Ades}{2006}]{LuAdes2006}
\begin{barticle}[mr]
\bauthor{\bsnm{Lu},~\bfnm{Guobing}\binits{G.}} \AND
\bauthor{\bsnm{Ades},~\bfnm{A.~E.}\binits{A.~E.}}
(\byear{2006}).
\btitle{Assessing evidence inconsistency in mixed treatment comparisons}.
\bjournal{J. Amer. Statist. Assoc.}
\bvolume{101}
\bpages{447--459}.
\bid{doi={10.1198/016214505000001302}, issn={0162-1459}, mr={2256166}}
\end{barticle}
%
\bptok{imsref}%
\endbibitem

\bibitem[\protect\citeauthoryear{Lunn et~al.}{2009}]{LunnEtAl2009}
\begin{barticle}[mr]
\bauthor{\bsnm{Lunn},~\bfnm{David}\binits{D.}},
\bauthor{\bsnm{Spiegelhalter},~\bfnm{David}\binits{D.}},
\bauthor{\bsnm{Thomas},~\bfnm{Andrew}\binits{A.}} \AND
\bauthor{\bsnm{Best},~\bfnm{Nicky}\binits{N.}}
(\byear{2009}).
\btitle{The BUGS project: Evolution, critique and future directions}.
\bjournal{Stat. Med.}
\bvolume{28}
\bpages{3049--3067}.
\bid{doi={10.1002/sim.3680}, issn={0277-6715}, mr={2750401}}
\end{barticle}
\bptok{imsref}%
\endbibitem

\bibitem[\protect\citeauthoryear{McDonald et~al.}{2014}]{McDonaldEtAl2013}
\begin{barticle}[author]
\bauthor{\bsnm{McDonald},~\bfnm{Scott~A.}\binits{S.~A.}},
\bauthor{\bsnm{Presanis},~\bfnm{Anne~M.}\binits{A.~M.}},
\bauthor{\bsnm{De~Angelis},~\bfnm{Daniela}\binits{D.}},
\bauthor{\bsnm{van~der Hoek},~\bfnm{Wim}\binits{W.}},
\bauthor{\bsnm{Hooiveld},~\bfnm{Mariette}\binits{M.}},
\bauthor{\bsnm{Donker},~\bfnm{G{\'{e}}}\binits{G.}} \AND
\bauthor{\bsnm{Kretzschmar},~\bfnm{Mirjam~E.}\binits{M.~E.}}
(\byear{2014}).
\btitle{An evidence synthesis approach to estimating the incidence of seasonal influenza in the Netherlands}.
\bjournal{Influenza Other Respi. Viruses}
\bvolume{8}
\bpages{33--41}.
\end{barticle}
%
\bptok{imsref}%
\endbibitem

\bibitem[\protect\citeauthoryear{Miller et~al.}{2009}]{MillerEtAl2009}
\begin{barticle}[pbm]
\bauthor{\bsnm{Miller},~\bfnm{Mark~A.}\binits{M.~A.}},
\bauthor{\bsnm{Viboud},~\bfnm{Cecile}\binits{C.}},
\bauthor{\bsnm{Balinska},~\bfnm{Marta}\binits{M.}} \AND
\bauthor{\bsnm{Simonsen},~\bfnm{Lone}\binits{L.}}
(\byear{2009}).
\btitle{The signature features of influenza pandemics--implications for policy}.
\bjournal{N. Engl. J. Med.}
\bvolume{360}
\bpages{2595--2598}.
\bid{doi={10.1056/NEJMp0903906}, issn={1533-4406}, pii={NEJMp0903906}, pmid={19423872}}
\end{barticle}
\bptok{imsref}%
\endbibitem

\bibitem[\protect\citeauthoryear{Miller et~al.}{2010}]{MillerEtAl2010}
\begin{barticle}[author]
\bauthor{\bsnm{Miller},~\bfnm{Elizabeth}\binits{E.}},
\bauthor{\bsnm{Hoschler},~\bfnm{Katja}\binits{K.}},
\bauthor{\bsnm{Hardelid},~\bfnm{Pia}\binits{P.}},
\bauthor{\bsnm{Stanford},~\bfnm{Elaine}\binits{E.}},
\bauthor{\bsnm{Andrews},~\bfnm{Nick}\binits{N.}} \AND
\bauthor{\bsnm{Zambon},~\bfnm{Maria}\binits{M.}}
(\byear{2010}).
\btitle{{Incidence of 2009 pandemic influenza A/H1N1 infection in England: A cross-sectional serological study}}.
\bjournal{The Lancet}
\bvolume{375}
\bpages{1100--1108}.
\end{barticle}
%
\bptok{imsref}%
\endbibitem

\bibitem[\protect\citeauthoryear{O'Hagan}{2003}]{OHagan2003}
\begin{bincollection}[mr]
\bauthor{\bsnm{O'Hagan},~\bfnm{Anthony}\binits{A.}}
(\byear{2003}).
\btitle{H{SSS} model criticism}.
In \bbooktitle{Highly Structured Stochastic Systems}.
\bseries{Oxford Statist. Sci. Ser.}
\bvolume{27}
\bpages{423--453}.
\bpublisher{Oxford Univ. Press},
\blocation{Oxford}.
\bid{mr={2082418}}
\bptnote{check related}%
\end{bincollection}
%
\bptok{imsref}%
\endbibitem

\bibitem[\protect\citeauthoryear{Pebody et~al.}{2010}]{PebodyEtAl2010}
\begin{barticle}[author]
\bauthor{\bsnm{Pebody},~\bfnm{R.~G.}\binits{R.~G.}},
\bauthor{\bsnm{McLean},~\bfnm{E.}\binits{E.}},
\bauthor{\bsnm{Zhao},~\bfnm{H.}\binits{H.}},
\bauthor{\bsnm{Cleary},~\bfnm{P.}\binits{P.}},
\bauthor{\bsnm{Bracebridge},~\bfnm{S.}\binits{S.}},
\bauthor{\bsnm{Foster},~\bfnm{K.}\binits{K.}},
\bauthor{\bsnm{Charlett},~\bfnm{A.}\binits{A.}},
\bauthor{\bsnm{Hardelid},~\bfnm{P.}\binits{P.}},
\bauthor{\bsnm{Waight},~\bfnm{P.}\binits{P.}},
\bauthor{\bsnm{Ellis},~\bfnm{J.}\binits{J.}},
\bauthor{\bsnm{Bermingham},~\bfnm{A.}\binits{A.}},
\bauthor{\bsnm{Zambon},~\bfnm{M.}\binits{M.}},
\bauthor{\bsnm{Evans},~\bfnm{B.}\binits{B.}},
\bauthor{\bsnm{Salmon},~\bfnm{R.}\binits{R.}},
\bauthor{\bsnm{McMenamin},~\bfnm{J.}\binits{J.}},
\bauthor{\bsnm{Smyth},~\bfnm{B.}\binits{B.}},
\bauthor{\bsnm{Catchpole},~\bfnm{M.}\binits{M.}} \AND
\bauthor{\bsnm{Watson},~\bfnm{Jm}\binits{J.}}
(\byear{2010}).
\btitle{{Pandemic influenza A(H1N1) 2009 and mortality in the United Kingdom: Risk factors for death, April 2009 to March 2010.}}
\bjournal{Euro Surveillance}
\bvolume{15}.
\end{barticle}
%
\bptok{imsref}%
\endbibitem

\bibitem[\protect\citeauthoryear{Presanis et~al.}{2008}]{PresanisEtAl2008}
\begin{barticle}[mr]
\bauthor{\bsnm{Presanis},~\bfnm{A.~M.}\binits{A.~M.}},
\bauthor{\bsnm{De Angelis},~\bfnm{D.}\binits{D.}},
\bauthor{\bsnm{Spiegelhalter},~\bfnm{D.~J.}\binits{D.~J.}},
\bauthor{\bsnm{Seaman},~\bfnm{S.}\binits{S.}},
\bauthor{\bsnm{Goubar},~\bfnm{A.}\binits{A.}} \AND
\bauthor{\bsnm{Ades},~\bfnm{A.~E.}\binits{A.~E.}}
(\byear{2008}).
\btitle{Conflicting evidence in a {B}ayesian synthesis of surveillance data to estimate human immunodeficiency virus prevalence}.
\bjournal{J. Roy. Statist. Soc. Ser. A}
\bvolume{171}
\bpages{915--937}.
\bid{doi={10.1111/j.1467-985X.2008.00543.x}, issn={0964-1998}, mr={2530293}}
\end{barticle}
\bptok{imsref}%
\endbibitem

\bibitem[\protect\citeauthoryear{Presanis et~al.}{2009}]{PresanisEtAl2009}
\begin{barticle}[pbm]
\bauthor{\bsnm{Presanis},~\bfnm{Anne~M.}\binits{A.~M.}},
\bauthor{\bsnm{De Angelis},~\bfnm{Daniela}\binits{D.}},
\bauthor{\bsnm{{New York City Swine Flu Investigation Team}}},
\bauthor{\bsnm{Hagy},~\bfnm{Angela}\binits{A.}},
\bauthor{\bsnm{Reed},~\bfnm{Carrie}\binits{C.}},
\bauthor{\bsnm{Riley},~\bfnm{Steven}\binits{S.}},
\bauthor{\bsnm{Cooper},~\bfnm{Ben~S.}\binits{B.~S.}},
\bauthor{\bsnm{Finelli},~\bfnm{Lyn}\binits{L.}},
\bauthor{\bsnm{Biedrzycki},~\bfnm{Paul}\binits{P.}} \AND
\bauthor{\bsnm{Lipsitch},~\bfnm{Marc}\binits{M.}}
(\byear{2009}).
\btitle{The severity of pandemic H1N1 influenza in the United States, from April to July 2009: A Bayesian analysis}.
\bjournal{PLoS Med.}
\bvolume{6}
\bpages{e1000207}.
\bid{doi={10.1371/journal.pmed.1000207}, issn={1549-1676}, pmcid={2784967}, pmid={19997612}}
\end{barticle}
%
\bptok{imsref}%
\endbibitem

\bibitem[\protect\citeauthoryear{Presanis et~al.}{2011a}]{PresanisEtAl2011}
\begin{barticle}[pbm]
\bauthor{\bsnm{Presanis},~\bfnm{A.~M.}\binits{A.~M.}},
\bauthor{\bsnm{De Angelis},~\bfnm{D.}\binits{D.}},
\bauthor{\bsnm{Goubar},~\bfnm{A.}\binits{A.}},
\bauthor{\bsnm{Gill},~\bfnm{O.~N.}\binits{O.~N.}} \AND
\bauthor{\bsnm{Ades},~\bfnm{A.~E.}\binits{A.~E.}}
(\byear{2011}a).
\btitle{Bayesian evidence synthesis for a transmission dynamic model for HIV among men who have sex with men}.
\bjournal{Biostatistics}
\bvolume{12}
\bpages{666--681}.
\bid{doi={10.1093/biostatistics/kxr006}, issn={1468-4357}, pii={kxr006}, pmcid={3169669}, pmid={21525422}}
\end{barticle}
\bptok{imsref}%
\endbibitem

\bibitem[\protect\citeauthoryear{Presanis et~al.}{2011b}]{PresanisEtAl2011a}
\begin{barticle}[author]
\bauthor{\bsnm{Presanis},~\bfnm{A.~M.}\binits{A.~M.}},
\bauthor{\bsnm{Pebody},~\bfnm{R.~G.}\binits{R.~G.}},
\bauthor{\bsnm{Paterson},~\bfnm{B.~J.}\binits{B.~J.}},
\bauthor{\bsnm{Tom},~\bfnm{B.~D.~M.}\binits{B.~D.~M.}},
\bauthor{\bsnm{Birrell},~\bfnm{P.~J.}\binits{P.~J.}},
\bauthor{\bsnm{Charlett},~\bfnm{A.}\binits{A.}},
\bauthor{\bsnm{Lipsitch},~\bfnm{M.}\binits{M.}} \AND
\bauthor{\bsnm{De~Angelis},~\bfnm{D.}\binits{D.}}
(\byear{2011}b).
\btitle{{Changes in severity of 2009 pandemic A/H1N1 influenza in England: A Bayesian evidence synthesis}}.
\bjournal{BMJ}
\bvolume{343}.
\end{barticle}
\bptok{imsref}%
\endbibitem

\bibitem[\protect\citeauthoryear{Presanis et~al.}{2013}]{PresanisEtAl2013}
\begin{barticle}[mr]
\bauthor{\bsnm{Presanis},~\bfnm{Anne~M.}\binits{A.~M.}},
\bauthor{\bsnm{Ohlssen},~\bfnm{David}\binits{D.}},
\bauthor{\bsnm{Spiegelhalter},~\bfnm{David~J.}\binits{D.~J.}} \AND
\bauthor{\bsnm{De Angelis},~\bfnm{Daniela}\binits{D.}}
(\byear{2013}).
\btitle{Conflict diagnostics in directed acyclic graphs, with applications in {B}ayesian evidence synthesis}.
\bjournal{Statist. Sci.}
\bvolume{28}
\bpages{376--397}.
\bid{doi={10.1214/13-STS426}, issn={0883-4237}, mr={3135538}}
\end{barticle}
%
\bptok{imsref}%
\endbibitem\



\bibitem[\protect\citeauthoryear{Presanis et~al.}{2014}]{PresanisEtAlSupp2013}
\begin{bmisc}[author]
\bauthor{\bsnm{Presanis},~\bfnm{A.~M.}\binits{A.~M.}},
\bauthor{\bsnm{Pebody},~\bfnm{R.~G.}\binits{R.~G.}},
\bauthor{\bsnm{Birrell},~\bfnm{P.~J.}\binits{P.~J.}},
\bauthor{\bsnm{Tom},~\bfnm{B.~D.~M.}\binits{B.~D.~M.}},
\bauthor{\bsnm{Green},~\bfnm{H.}\binits{H.}},
\bauthor{\bsnm{Durnell},~\bfnm{H.}\binits{H.}},
\bauthor{\bsnm{Fleming},~\bfnm{D.}\binits{D.}} \AND
\bauthor{\bsnm{De~Angelis},~\bfnm{D.}\binits{D.}}
(\byear{2014}).
\bhowpublished{Supplement to ``Synthesising evidence to estimate pandemic (2009) A/H1N1
influenza severity in 2009--2011.''
DOI:\doiurl{10.1214/14-AOAS775SUPP}}.
%
\end{bmisc}
\bptok{imsref}
\endbibitem


\bibitem[\protect\citeauthoryear{Reed et~al.}{2009}]{ReedEtAl2009}
\begin{barticle}[pbm]
\bauthor{\bsnm{Reed},~\bfnm{Carrie}\binits{C.}},
\bauthor{\bsnm{Angulo},~\bfnm{Frederick~J.}\binits{F.~J.}},
\bauthor{\bsnm{Swerdlow},~\bfnm{David~L.}\binits{D.~L.}},
\bauthor{\bsnm{Lipsitch},~\bfnm{Marc}\binits{M.}},
\bauthor{\bsnm{Meltzer},~\bfnm{Martin~I.}\binits{M.~I.}},
\bauthor{\bsnm{Jernigan},~\bfnm{Daniel}\binits{D.}} \AND
\bauthor{\bsnm{Finelli},~\bfnm{Lyn}\binits{L.}}
(\byear{2009}).
\btitle{Estimates of the prevalence of pandemic (H1N1) 2009, United States, April--July 2009}.
\bjournal{Emerging Infect. Dis.}
\bvolume{15}
\bpages{2004--2007}.
\bid{doi={10.3201/eid1512.091413}, issn={1080-6059}, pmcid={3375879}, pmid={19961687}}
\end{barticle}
\bptok{imsref}%
\endbibitem

\bibitem[\protect\citeauthoryear{Shubin et~al.}{2013}]{ShubinEtAl2013}
\begin{barticle}[author]
\bauthor{\bsnm{Shubin},~\bfnm{M.}\binits{M.}},
\bauthor{\bsnm{Virtanen},~\bfnm{M.}\binits{M.}},
\bauthor{\bsnm{Toikkanen},~\bfnm{S.}\binits{S.}},
\bauthor{\bsnm{Lyytik{\"{a}}inen},~\bfnm{O.}\binits{O.}} \AND
\bauthor{\bsnm{Auranen},~\bfnm{K.}\binits{K.}}
(\byear{2013}).
\btitle{Estimating the burden of {A(H1N1})pdm09 influenza in Finland during two seasons}.
\bjournal{Epidemiology~\& Infection}
\bvolume{142}
\bpages{964--974}.
\end{barticle}
%
\bptok{imsref}%
\endbibitem

\bibitem[\protect\citeauthoryear{Spiegelhalter, Abrams and Myles}{2004}]{SpiegelhalterEtAl2004}
\begin{bbook}[author]
\bauthor{\bsnm{Spiegelhalter},~\bfnm{David~J.}\binits{D.~J.}},
\bauthor{\bsnm{Abrams},~\bfnm{Keith~R.}\binits{K.~R.}} \AND
\bauthor{\bsnm{Myles},~\bfnm{Jonathan~P.}\binits{J.~P.}}
(\byear{2004}).
\btitle{{Bayesian Approaches to Clinical Trials and Health-Care Evaluation (Statistics in Practice)}}.
\bpublisher{Wiley},
\blocation{New York}.
\end{bbook}
\bptok{imsref}%
\endbibitem

\bibitem[\protect\citeauthoryear{Sweeting et~al.}{2008}]{SweetingEtAl2008}
\begin{barticle}[pbm]
\bauthor{\bsnm{Sweeting},~\bfnm{M.~J.}\binits{M.~J.}},
\bauthor{\bsnm{De Angelis},~\bfnm{D.}\binits{D.}},
\bauthor{\bsnm{Hickman},~\bfnm{M.}\binits{M.}} \AND
\bauthor{\bsnm{Ades},~\bfnm{A.~E.}\binits{A.~E.}}
(\byear{2008}).
\btitle{Estimating hepatitis C prevalence in England and Wales by synthesizing evidence from multiple data sources. Assessing data conflict and model fit}.
\bjournal{Biostatistics}
\bvolume{9}
\bpages{715--734}.
\bid{doi={10.1093/biostatistics/kxn004}, issn={1468-4357}, pii={kxn004}, pmid={18349037}}
\end{barticle}
%
\bptok{imsref}%
\endbibitem

\bibitem[\protect\citeauthoryear{Sypsa et~al.}{2011}]{SypsaEtAl2011}
\begin{barticle}[pbm]
\bauthor{\bsnm{Sypsa},~\bfnm{Vana}\binits{V.}},
\bauthor{\bsnm{Bonovas},~\bfnm{Stefanos}\binits{S.}},
\bauthor{\bsnm{Tsiodras},~\bfnm{Sotirios}\binits{S.}},
\bauthor{\bsnm{Baka},~\bfnm{Agoritsa}\binits{A.}},
\bauthor{\bsnm{Efstathiou},~\bfnm{Panos}\binits{P.}},
\bauthor{\bsnm{Malliori},~\bfnm{Meni}\binits{M.}},
\bauthor{\bsnm{Panagiotopoulos},~\bfnm{Takis}\binits{T.}},
\bauthor{\bsnm{Nikolakopoulos},~\bfnm{Ilias}\binits{I.}} \AND
\bauthor{\bsnm{Hatzakis},~\bfnm{Angelos}\binits{A.}}
(\byear{2011}).
\btitle{Estimating the disease burden of 2009 pandemic influenza A(H1N1) from surveillance and household surveys in Greece}.
\bjournal{PLoS ONE}
\bvolume{6}
\bpages{e20593}.
\bid{doi={10.1371/journal.pone.0020593}, issn={1932-6203}, pii={PONE-D-11-03363}, pmcid={3111416}, pmid={21694769}}
\end{barticle}
\bptok{imsref}%
\endbibitem

\bibitem[\protect\citeauthoryear{Truelove et~al.}{2011}]{TrueloveEtAl2011}
\begin{barticle}[pbm]
\bauthor{\bsnm{Truelove},~\bfnm{Shaun~A.}\binits{S.~A.}},
\bauthor{\bsnm{Chitnis},~\bfnm{Amit~S.}\binits{A.~S.}},
\bauthor{\bsnm{Heffernan},~\bfnm{Richard~T.}\binits{R.~T.}},
\bauthor{\bsnm{Karon},~\bfnm{Amy~E.}\binits{A.~E.}},
\bauthor{\bsnm{Haupt},~\bfnm{Thomas~E.}\binits{T.~E.}} \AND
\bauthor{\bsnm{Davis},~\bfnm{Jeffrey~P.}\binits{J.~P.}}
(\byear{2011}).
\btitle{Comparison of patients hospitalized with pandemic 2009 influenza A(H1N1) virus infection during the first two pandemic waves in Wisconsin}.
\bjournal{J. Infect. Dis.}
\bvolume{203}
\bpages{828--837}.
\bid{doi={10.1093/infdis/jiq117}, issn={1537-6613}, pii={jiq117}, pmcid={3071126}, pmid={21278213}}
\end{barticle}
%
\bptok{imsref}%
\endbibitem

\bibitem[\protect\citeauthoryear{Van~Kerkhove et~al.}{2011}]{VanKerkhoveEtAl2011}
\begin{barticle}[author]
\bauthor{\bsnm{Van~Kerkhove},~\bfnm{Maria~D.}\binits{M.~D.}},
\bauthor{\bsnm{Vandemaele},~\bfnm{Katelijn~A.~H.}\binits{K.~A.~H.}},
\bauthor{\bsnm{Shinde},~\bfnm{Vivek}\binits{V.}},
\bauthor{\bsnm{Jaramillo-Gutierrez},~\bfnm{Giovanna}\binits{G.}},
\bauthor{\bsnm{Koukounari},~\bfnm{Artemis}\binits{A.}},
\bauthor{\bsnm{Donnelly},~\bfnm{Christl~A.}\binits{C.~A.}},
\bauthor{\bsnm{Carlino},~\bfnm{Luis~O.}\binits{L.~O.}},
\bauthor{\bsnm{Owen},~\bfnm{Rhonda}\binits{R.}},
\bauthor{\bsnm{Paterson},~\bfnm{Beverly}\binits{B.}},
\bauthor{\bsnm{Pelletier},~\bfnm{Louise}\binits{L.}},
\bauthor{\bsnm{Vachon},~\bfnm{Julie}\binits{J.}},
\bauthor{\bsnm{Gonzalez},~\bfnm{Claudia}\binits{C.}},
\bauthor{\bsnm{Hongjie},~\bfnm{Yu}\binits{Y.}},
\bauthor{\bsnm{Zijian},~\bfnm{Feng}\binits{F.}},
\bauthor{\bsnm{Chuang},~\bfnm{Shuk~K.}\binits{S.~K.}},
\bauthor{\bsnm{Au},~\bfnm{Albert}\binits{A.}},
\bauthor{\bsnm{Buda},~\bfnm{Silke}\binits{S.}},
\bauthor{\bsnm{Krause},~\bfnm{Gerard}\binits{G.}},
\bauthor{\bsnm{Haas},~\bfnm{Walter}\binits{W.}},
\bauthor{\bsnm{Bonmarin},~\bfnm{Isabelle}\binits{I.}},
\bauthor{\bsnm{Taniguichi},~\bfnm{Kiyosu}\binits{K.}},
\bauthor{\bsnm{Nakajima},~\bfnm{Kensuke}\binits{K.}},
\bauthor{\bsnm{Shobayashi},~\bfnm{Tokuaki}\binits{T.}},
\bauthor{\bsnm{Takayama},~\bfnm{Yoshihiro}\binits{Y.}},
\bauthor{\bsnm{Sunagawa},~\bfnm{Tomi}\binits{T.}},
\bauthor{\bsnm{Heraud},~\bfnm{Jean~M.}\binits{J.~M.}},
\bauthor{\bsnm{Orelle},~\bfnm{Arnaud}\binits{A.}},
\bauthor{\bsnm{Palacios},~\bfnm{Ethel}\binits{E.}},
\bauthor{\bsnm{van~der Sande},~\bfnm{Marianne~A.~B.}\binits{M.~A.~B.}},
\bauthor{\bsnm{Wielders},~\bfnm{C.~C.~H.~Lieke}\binits{C.~C.~H.~L.}},
\bauthor{\bsnm{Hunt},~\bfnm{Darren}\binits{D.}},
\bauthor{\bsnm{Cutter},~\bfnm{Jeffrey}\binits{J.}},
\bauthor{\bsnm{Lee},~\bfnm{Vernon~J.}\binits{V.~J.}},
\bauthor{\bsnm{Thomas},~\bfnm{Juno}\binits{J.}},
\bauthor{\bsnm{Santa-Olalla},~\bfnm{Patricia}\binits{P.}},
\bauthor{\bsnm{Sierra-Moros},~\bfnm{Maria~J.}\binits{M.~J.}},
\bauthor{\bsnm{Hanshaoworakul},~\bfnm{Wanna}\binits{W.}},
\bauthor{\bsnm{Ungchusak},~\bfnm{Kumnuan}\binits{K.}},
\bauthor{\bsnm{Pebody},~\bfnm{Richard}\binits{R.}},
\bauthor{\bsnm{Jain},~\bfnm{Seema}\binits{S.}},
\bauthor{\bsnm{Mounts},~\bfnm{Anthony~W.}\binits{A.~W.}} \AND
\bauthor{\bsnm{on~behalf~of~the~WHO~Working~Group~for~Risk~Factors~forSevere H1N1pdm~Infection}}
(\byear{2011}).
\btitle{Risk factors for severe outcomes following 2009 influenza~A({H1N1}) infection: A~global pooled analysis}.
\bjournal{PLoS Med}
\bvolume{8}
\bpages{e1001053+}.
\end{barticle}
%
\bptok{imsref}%
\endbibitem

\bibitem[\protect\citeauthoryear{Welton and Ades}{2005}]{WeltonAdes2005}
\begin{barticle}[mr]
\bauthor{\bsnm{Welton},~\bfnm{N.~J.}\binits{N.~J.}} \AND
\bauthor{\bsnm{Ades},~\bfnm{A.~E.}\binits{A.~E.}}
(\byear{2005}).
\btitle{A model of toxoplasmosis incidence in the UK: Evidence synthesis and consistency of evidence}.
\bjournal{J. R. Stat. Soc. Ser. C. Appl. Stat.}
\bvolume{54}
\bpages{385--404}.
\bid{doi={10.1111/j.1467-9876.2005.00490.x}, issn={0035-9254}, mr={2135881}}
\end{barticle}
%
\bptok{imsref}%
\endbibitem

\bibitem[\protect\citeauthoryear{Wielders et~al.}{2012}]{WieldersEtAl2010}
\begin{barticle}[author]
\bauthor{\bsnm{Wielders},~\bfnm{C.~C.~H.}\binits{C.~C.~H.}},
\bauthor{\bparticle{van} \bsnm{Lier},~\bfnm{E.~A.}\binits{E.~A.}},
\bauthor{\bsnm{van't Klooster},~\bfnm{T.~M.}\binits{T.~M.}},
\bauthor{\bparticle{van} \bsnm{Gageldonk-Lafeber},~\bfnm{A.~B.}\binits{A.~B.}},
\bauthor{\bsnm{van~den Wijngaard},~\bfnm{C.~C.}\binits{C.~C.}},
\bauthor{\bsnm{Haagsma},~\bfnm{J.~A.}\binits{J.~A.}},
\bauthor{\bsnm{Donker},~\bfnm{G.~A.}\binits{G.~A.}},
\bauthor{\bsnm{Meijer},~\bfnm{A.}\binits{A.}},
\bauthor{\bsnm{van~der Hoek},~\bfnm{W.}\binits{W.}},
\bauthor{\bsnm{Lugn{\'{e}}r},~\bfnm{A.~K.}\binits{A.~K.}},
\bauthor{\bsnm{Kretzschmar},~\bfnm{M.~E.~E.}\binits{M.~E.~E.}} \AND
\bauthor{\bsnm{van~der Sande},~\bfnm{M.~A.~B.}\binits{M.~A.~B.}}
(\byear{2012}).
\btitle{The burden of 2009 pandemic influenza {A(H1N1}) in the Netherlands}.
\bjournal{The European Journal of Public Health}
\bvolume{22}
\bpages{150--157}.
\end{barticle}
%
\bptok{imsref}%
\endbibitem

\bibitem[\protect\citeauthoryear{Wilson and Baker}{2009}]{WilsonBaker2009}
\begin{barticle}[pbm]
\bauthor{\bsnm{Wilson},~\bfnm{N.}\binits{N.}} \AND
\bauthor{\bsnm{Baker},~\bfnm{M.~G.}\binits{M.~G.}}
(\byear{2009}).
\btitle{The emerging influenza pandemic: Estimating the case fatality ratio}.
\bjournal{Euro Surveill.}
\bvolume{14}.
\bid{issn={1560-7917}, pmid={19573509}}
\end{barticle}
\bptok{imsref}%
\endbibitem

\bibitem[\protect\citeauthoryear{Wong et~al.}{2013}]{WongEtAl2013}
\begin{barticle}[author]
\bauthor{\bsnm{Wong},~\bfnm{Jessica~Y.}\binits{J.~Y.}},
\bauthor{\bsnm{Kelly},~\bfnm{Heath}\binits{H.}},
\bauthor{\bsnm{Ip},~\bfnm{Dennis~K.}\binits{D.~K.}},
\bauthor{\bsnm{Wu},~\bfnm{Joseph~T.}\binits{J.~T.}},
\bauthor{\bsnm{Leung},~\bfnm{Gabriel~M.}\binits{G.~M.}} \AND
\bauthor{\bsnm{Cowling},~\bfnm{Benjamin~J.}\binits{B.~J.}}
(\byear{2013}).
\btitle{Case fatality risk of influenza A({H1N1pdm09}): A systematic review.}
\bjournal{Epidemiology (Cambridge, Mass.)}
\bvolume{24}
\bpages{830--841}.
\end{barticle}
%
\bptok{imsref}%
\endbibitem

\bibitem[\protect\citeauthoryear{Wu et~al.}{2010}]{WuEtAl2010}
\begin{barticle}[author]
\bauthor{\bsnm{Wu},~\bfnm{Joseph~T.}\binits{J.~T.}},
\bauthor{\bsnm{Ma},~\bfnm{Edward~S.~K.}\binits{E.~S.~K.}},
\bauthor{\bsnm{Lee},~\bfnm{Cheuk~K.}\binits{C.~K.}},
\bauthor{\bsnm{Chu},~\bfnm{Daniel~K.~W.}\binits{D.~K.~W.}},
\bauthor{\bsnm{Ho},~\bfnm{Po-Lai}\binits{P.-L.}},
\bauthor{\bsnm{Shen},~\bfnm{Angela~L.}\binits{A.~L.}},
\bauthor{\bsnm{Ho},~\bfnm{Andrew}\binits{A.}},
\bauthor{\bsnm{Hung},~\bfnm{Ivan~F.~N.}\binits{I.~F.~N.}},
\bauthor{\bsnm{Riley},~\bfnm{Steven}\binits{S.}},
\bauthor{\bsnm{Ho},~\bfnm{Lai~M.}\binits{L.~M.}},
\bauthor{\bsnm{Lin},~\bfnm{Che~K.}\binits{C.~K.}},
\bauthor{\bsnm{Tsang},~\bfnm{Thomas}\binits{T.}},
\bauthor{\bsnm{Lo},~\bfnm{Su-Vui}\binits{S.-V.}},
\bauthor{\bsnm{Lau},~\bfnm{Yu-Lung}\binits{Y.-L.}},
\bauthor{\bsnm{Leung},~\bfnm{Gabriel~M.}\binits{G.~M.}},
\bauthor{\bsnm{Cowling},~\bfnm{Benjamin~J.}\binits{B.~J.}} \AND
\bauthor{\bsnm{Peiris},~\bfnm{J.~S.~Malik}\binits{J.~S.~M.}}
(\byear{2010}).
\btitle{The infection attack rate and severity of 2009 pandemic {H1N1} influenza in Hong Kong}.
\bjournal{Clinical Infectious Diseases}
\bvolume{51}
\bpages{1184--1191}.
\end{barticle}
%
\bptok{imsref}%
\endbibitem\

\bibitem[\protect\citeauthoryear{Yang et~al.}{2011}]{YangEtAl2011}
\begin{barticle}[author]
\bauthor{\bsnm{Yang},~\bfnm{Ji-Rong}\binits{J.-R.}},
\bauthor{\bsnm{Huang},~\bfnm{Yuan-Pin}\binits{Y.-P.}},
\bauthor{\bsnm{Chang},~\bfnm{Feng-Yee}\binits{F.-Y.}},
\bauthor{\bsnm{Hsu},~\bfnm{Li-Ching}\binits{L.-C.}},
\bauthor{\bsnm{Lin},~\bfnm{Yu-Cheng}\binits{Y.-C.}},
\bauthor{\bsnm{Su},~\bfnm{Chun-Hui}\binits{C.-H.}},
\bauthor{\bsnm{Chen},~\bfnm{Pei-Jer}\binits{P.-J.}},
\bauthor{\bsnm{Wu},~\bfnm{Ho-Sheng}\binits{H.-S.}} \AND
\bauthor{\bsnm{Liu},~\bfnm{Ming-Tsan}\binits{M.-T.}}
(\byear{2011}).
\btitle{New variants and age shift to high fatality groups contribute to severe successive waves in the 2009 influenza pandemic in Taiwan}.
\bjournal{PLoS ONE}
\bvolume{6}
\bpages{e28288+}.
\end{barticle}
%
\bptok{imsref}%
\endbibitem
\end{thebibliography}
\end{document}